\documentclass{article}

\pdfoutput=1

\usepackage[nonatbib, preprint]{neurips_data_2024}

\usepackage[utf8]{inputenc} 
\usepackage[T1]{fontenc}    
\usepackage{hyperref}       
\usepackage{url}            
\usepackage{booktabs}       
\usepackage{amsfonts}       
\usepackage{nicefrac}       
\usepackage{microtype}      
\usepackage{xcolor}         
\usepackage{tcolorbox}
\usepackage{subcaption}
\usepackage{amsmath}
\usepackage{amssymb}
\usepackage{bm}
\usepackage{multirow}
\usepackage{makecell}
\usepackage{upgreek}
\usepackage{listings}
\usepackage{adjustbox}

\definecolor{codegreen}{rgb}{0,0.6,0}
\definecolor{codegray}{rgb}{0.5,0.5,0.5}
\definecolor{codepurple}{rgb}{0.58,0,0.82}
\definecolor{backcolour}{rgb}{0.95,0.95,0.95}
\lstdefinestyle{mystyle}{
    backgroundcolor=\color{backcolour},   
    commentstyle=\color{codegreen},
    keywordstyle=\color{magenta},
    numberstyle=\tiny\color{codegray},
    stringstyle=\color{codepurple},
    basicstyle=\tiny,
    breakatwhitespace=false,         
    breaklines=true,                 
    captionpos=b,                    
    keepspaces=true,                 
    numbers=none,                    
    numbersep=3pt,                  
    showspaces=false,                
    showstringspaces=false,
    showtabs=false,                  
    tabsize=2
}

\usepackage{tikz}
\usepackage{pgfplots}
\tikzset{every picture/.style={/utils/exec={\sffamily}}}
\usepackage{environ}
\usepackage{circuitikz}
\pgfplotsset{compat = newest}
\usetikzlibrary{arrows,positioning} 
\usetikzlibrary{decorations.text}
\usetikzlibrary{decorations.pathreplacing, calc}
\usetikzlibrary{arrows.meta}

\pgfplotsset{
compat=1.11,
legend image code/.code={
\draw[mark repeat=2,mark phase=2]
plot coordinates {
(0cm,0cm)
(0.15cm,0cm)        
(0.3cm,0cm)         
};%
}
}

\makeatletter
\tikzset{
    >=stealth',
    port/.style = {circle, draw, align=center, minimum height=1mm},
    op/.style={
           rectangle,
           rounded corners,
           draw=black, thick,
           text width=3.5em,
           minimum height=1em,
           text centered},
    conv/.style={
           rectangle,
           rounded corners,
           draw=black, thick,
           text width=3.5em,
           minimum height=1em,
           text centered},
    data/.style={
           rectangle,
           draw=black, thick,
           minimum height=1em,
           text centered},
    area/.style={
           rectangle,
           draw=black, thick,
           minimum height=1em},
    connect/.style={
           ->,
           thick,
           shorten <=2pt,
           shorten >=2pt,},
    database segment style/.style={draw},
    database/.style={
        path picture={
            \path [database segment]
                (-\db@r,-0.5*\db@sh) 
                -- ++(0,-1*\db@sh) 
                arc [start angle=180, end angle=360,
                    x radius=\db@r, y radius=\db@ar*\db@r]
                -- ++(0,1*\db@sh)
                arc [start angle=360, end angle=180,
                    x radius=\db@r, y radius=\db@ar*\db@r];
            \path [database segment style]
                (-\db@r,0.5*\db@sh) 
                -- ++(0,-1*\db@sh) 
                arc [start angle=180, end angle=360,
                    x radius=\db@r, y radius=\db@ar*\db@r]
                -- ++(0,1*\db@sh)
                arc [start angle=360, end angle=180,
                    x radius=\db@r, y radius=\db@ar*\db@r];
            \path [database segment style]
                (-\db@r,1.5*\db@sh) 
                -- ++(0,-1*\db@sh) 
                arc [start angle=180, end angle=360,
                    x radius=\db@r, y radius=\db@ar*\db@r]
                -- ++(0,1*\db@sh)
                arc [start angle=360, end angle=180,
                    x radius=\db@r, y radius=\db@ar*\db@r];
            \path [database segment style]
                (0, 1.5*\db@sh) circle [x radius=\db@r, y radius=\db@ar*\db@r];
        },
        minimum width=2*\db@r + \pgflinewidth,
        minimum height=3*\db@sh + 2*\db@ar*\db@r + \pgflinewidth,
    },
    database segment height/.store in=\db@sh,
    database radius/.store in=\db@r,
    database aspect ratio/.store in=\db@ar,
    database segment height=0.1cm,
    database radius=0.25cm,
    database aspect ratio=0.35,
    database segment/.style={
        database segment style/.append style={#1}},
}
\makeatother

\newcommand{\method}{AICircuit}
\newcommand{\dataset}{AICircuit}

\title{AICircuit: A Multi-Level Dataset and Benchmark for AI-Driven Analog Integrated Circuit Design}

\author{%
  Asal Mehradfar$^{1}$ \quad
  Xuzhe Zhao$^{2}$ \quad
  Yue Niu$^{1}$ \quad
  Sara Babakniya$^{1}$ \quad
  Mahdi Alesheikh$^{2}$ \\
  \textbf{Hamidreza Aghasi}$^{2}$ \quad
  \textbf{Salman Avestimehr}$^{1}$ \\
  $^{1}$ University of Southern California \quad
  $^{2}$ University of California, Irvine
}

\begin{document}

\maketitle

\begin{abstract}

Analog and radio-frequency circuit design requires extensive exploration of both circuit topology and parameters to meet specific design criteria like power consumption and bandwidth. Designers must review state-of-the-art topology configurations in the literature and sweep various circuit parameters within each configuration. This design process is highly specialized and time-intensive, particularly as the number of circuit parameters increases and the circuit becomes more complex. Prior research has explored the potential of machine learning to enhance circuit design procedures. However, these studies primarily focus on simple circuits, overlooking the more practical and complex analog and radio-frequency systems. A major obstacle for bearing the power of machine learning in circuit design is the availability of a generic and diverse dataset, along with robust metrics, which are essential for thoroughly evaluating and improving machine learning algorithms in the analog and radio-frequency circuit domain. We present \method{}, a comprehensive multi-level dataset and benchmark for developing and evaluating ML algorithms in analog and radio-frequency circuit design. \method{} comprises seven commonly used basic circuits and two complex wireless transceiver systems composed of multiple circuit blocks, encompassing a wide array of design scenarios encountered in real-world applications. We extensively evaluate various ML algorithms on the dataset, revealing the potential of ML algorithms in learning the mapping from the design specifications to the desired circuit parameters. 
The data and codebase are open-sourced to advance the development of machine learning in the analog and radio-frequency circuit design domain in the following link: \url{https://github.com/AvestimehrResearchGroup/AICircuit}.
\end{abstract}

\section{Introduction}\label{sec:intro}

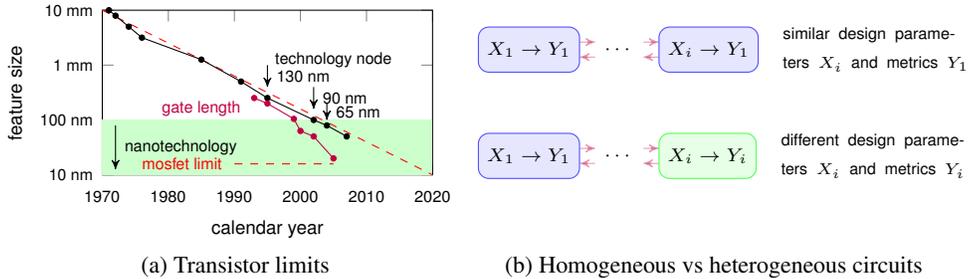
\begin{figure}[!htb]
\centering
\begin{subfigure}{.45\textwidth}
\begin{tikzpicture}
    \begin{axis}[
        width=.95\linewidth, height=.6\linewidth,
        xmin=0, xmax=5,
        xtick={0, 1, 2, 3, 4, 5},
        xticklabels={ 1970, 1980, 1990, 2000, 2010, 2020 },
        xticklabel style={ align=center,text width=14mm, outer sep=1mm },
        xlabel={\scriptsize calendar year},
        ymin=0, ymax=3,
        ylabel={\scriptsize feature size},
        ytick={0, 1, 2, 3},
        yticklabels={10 nm, 100 nm, 1 mm, 10 mm},
        ticklabel style={font=\tiny},
        legend style={at={(1.0,0.35)}},
    ]
        \fill [green!20]
        (0, 0) -- (0, 1) -- (5, 1) -- (5, 0) -- cycle;
        \draw[->] (.2, .9) -- node[right] {\tiny nanotechnology} (.2, .1);

        \addplot+[ 
            thin, red, dashed, mark=none
        ] 
        table [x=x, y=y, col sep=comma]{
            x,    y   
            0,    3
            5,    0
        };

        \addplot+[ 
            thin, black, mark=*, mark options={solid, scale=0.5}
        ] 
        table [x=x, y=y, col sep=comma]{
            x,    y   
            .1,   3
            .2,   2.9
            .4,   2.7
            .6,   2.5
            1.5,  2.1
            2.1,  1.7
            2.5,  1.4
            3.2,  1
            3.4,  .9
            3.7,  .7
        };
        \draw[->] (2.5, 2.0) -- node[right] {\tiny 130 nm} (2.5, 1.6);
        \draw[->] (3.2, 1.6) -- node[right] {\tiny 90 nm} (3.2, 1.2);
        \draw[->] (3.4, 1.3) -- node[right] {\tiny 65 nm} (3.4, 1.0);

        \addplot+[ 
            thin, purple, mark=*, mark options={solid, scale=0.5}
        ] 
        table [x=x, y=y, col sep=comma]{
            x,    y   
            2.3,  1.4
            2.5,  1.3
            2.9,  1.02
            3.0,  .8
            3.2,  .7
            3.5,  .3
        };
        \draw[-, red, dashed] (2, .2) -- node[left=.7cm] {\tiny mosfet limit} (3.5, .2);

        \node[fill=none, draw=none] at (axis cs:1.5,1.2){\tiny \textcolor{purple}{gate length}};
        \node[fill=none, draw=none] at (axis cs:3.5,2.1){\tiny \textcolor{black}{technology node}};
    \end{axis}
\end{tikzpicture}
\vspace{-5mm}
\caption{Transistor limits}
\label{fig:trend:transistor}
\end{subfigure}
\hspace{-1mm}
\begin{subfigure}{.45\textwidth}
\begin{tikzpicture}
    \node[minimum height=.6cm, rounded corners, fill=blue!10, draw=blue!70](hmB1){\scriptsize $X_1 \rightarrow Y_1$};
    \node[right=.2cm of hmB1, fill=none, draw=none](hmBi){\scriptsize $\cdots$};
    \node[right=.2cm of hmBi, minimum height=.6cm, rounded corners, fill=blue!10, draw=blue!70](hmBn){\scriptsize $X_i \rightarrow Y_1$};
    \node[right=.2cm of hmBn, text width=2.5cm]{\tiny similar design parameters $X_i$ and metrics $Y_1$};

    \draw[->, purple, opacity=.4] ($(hmB1.east)+(0mm, .1cm)$) -- +(.2cm, 0cm);
    \draw[<-, purple, opacity=.4] ($(hmB1.east)+(0mm, -.1cm)$) -- +(.2cm, 0cm);
    \draw[->, purple, opacity=.4] ($(hmBi.east)+(0mm, .1cm)$) -- +(.2cm, 0cm);
    \draw[<-, purple, opacity=.4] ($(hmBi.east)+(0mm, -.1cm)$) -- +(.2cm, 0cm);

    \node[below=.8cm of hmB1, minimum height=.6cm, rounded corners, fill=blue!10, draw=blue!70](htB1){\scriptsize $X_1 \rightarrow Y_1$};
    \node[right=.2cm of htB1, fill=none, draw=none](htBi){\scriptsize $\cdots$};
    \node[right=.2cm of htBi, minimum height=.6cm, rounded corners, fill=green!10, draw=green!70](htBn){\scriptsize $X_i \rightarrow Y_i$};
    \node[right=.2cm of htBn, text width=2.5cm]{\tiny different design parameters $X_i$ and metrics $Y_i$};

    \draw[->, purple, opacity=.4] ($(htB1.east)+(0mm, .1cm)$) -- +(.2cm, 0cm);
    \draw[<-, purple, opacity=.4] ($(htB1.east)+(0mm, -.1cm)$) -- +(.2cm, 0cm);
    \draw[->, purple, opacity=.4] ($(htBi.east)+(0mm, .1cm)$) -- +(.2cm, 0cm);
    \draw[<-, purple, opacity=.4] ($(htBi.east)+(0mm, -.1cm)$) -- +(.2cm, 0cm);
\end{tikzpicture}
\vspace{.3cm}
\caption{Homogeneous vs heterogeneous circuits}
\label{fig:trend:vol}
\end{subfigure}
\caption{\footnotesize The advances in analog and mm-wave circuit design. (a) The limits of transistor scaling predicted by Moore’s law; (b) Comparison between homogeneous and heterogeneous circuits.}
\label{fig:trend}
\end{figure}

By reaching the limits of Moore's Law \cite{IEEE97_Moore} in early 2020 (Figure \ref{fig:trend:transistor}), the exponential scaling of transistors, the core elements in circuit design, has become increasingly difficult, thereby slowing down the pace of advancements in semiconductor technologies. Unlike digital circuits, where scaling can often lead to straightforward performance improvements, analog and radio-frequency circuits should be custom-designed for emerging applications, e.g., mm-wave cellular communications, radar systems, and antenna systems, making their design both time-consuming and resource-intensive \cite{mmwave_celluar_system,mmwave_radar_system,mmwave_antenna_system,mmwave_application_review,Els21_Analog}. Moreover, the volume of the market for analog integrated circuits is surpassing all other sectors and readily it is the major driving force for the development of semiconductors. 
By transitioning into the analog domain, two major types of circuits are envisioned: i) homogeneous circuits with constituent sub-circuits with similar functions; and ii) heterogeneous circuits where the constituent elements behave differently and have non-identical performance metrics. 
Most mm-wave systems of interest are heterogeneous, and their constituent circuit blocks of various functionalities should be combined to achieve \textit{system level} performance metrics that meet the thresholds. 
It is challenging to obtain the optimal design points due to two fundamental reasons: 1) the performance of a circuit block, when operated individually, is different from when it is inside a system due to interaction among the circuit blocks; 2) various system-level metrics rely on the metrics of individual circuits in different ways, causing trade-off among the metrics. 


\begin{figure}[t]
    \centering
    \ctikzset{tripoles/mos style/arrows}
    \ctikzset{capacitors/scale=0.4}
    \ctikzset{resistors/scale=0.4}
    \ctikzset{transistors/scale=0.4, current arrow scale=24,}
    \ctikzset{inductors/scale=0.5}
    \begin{circuitikz}[american, cute inductors]
        \node[op, draw=none, fill=none, text width=1.5cm, minimum height=1cm](SIMULATION){\includegraphics[height=1.5cm, bb=0 0 200 200]{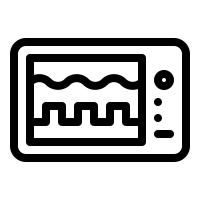}};
        \node[below=-4mm of SIMULATION](){\scriptsize verification};
        
        \node[op, right=5mm of SIMULATION, draw=black, fill=white, text width=1.5cm, minimum height=1cm](PERF){\tiny specifications};
        \node[below=1mm of PERF, xshift=.8cm, text width=3cm](){\baselineskip=6pt \tiny - bandwidth \\ - voltage gain, $\cdots$ \par};
        
        \node[op, left=3mm of SIMULATION, draw=black, fill=white, text width=2cm, minimum height=6mm](SWEEP){\scriptsize param sweeping};
        
        \node[op, left=5mm of SWEEP, draw=black, fill=white, text width=2cm, minimum height=6mm](TOPOLOGY){\scriptsize schematic design};
        
        \node[draw=black, fill=none, left=3mm of TOPOLOGY, yshift=0mm, minimum width=2cm, minimum height=1.2cm, rounded corners](COMPONENTS){};
        \node[below=1mm of COMPONENTS, xshift=1.1cm, text width=4cm](){\baselineskip=6pt \tiny circuit components \\ - transistors, $\cdots$ \par};
        
        \node[nmos, below=.5mm of COMPONENTS.north, xshift=-.6cm](NMOS){};
        \draw ($(NMOS.south) + (-2mm, -3mm)$) to [C, name=CAP] ++(0.5, 0);
        
        \draw ($(NMOS.east) + (5mm, 0mm)$) to [L, name=IND] ++(0.8, 0);

        \draw ($(CAP.east) + (6mm, 0mm)$) to [R, name=RES] ++(0.8, 0);

        \draw[->] ($(COMPONENTS.east) + (0mm, 2mm)$) to [out=45, in=135] ($(TOPOLOGY.west) + (0mm, 2mm)$);
        \draw[->] ($(TOPOLOGY.west) + (0mm, -2mm)$) to [out=-135, in=-45] ($(COMPONENTS.east) + (0mm, -2mm)$);
        \draw[->] (TOPOLOGY.east) to (SWEEP.west);
        \draw[->, shorten >=-6pt] (SWEEP.east) to (SIMULATION.west);
        \draw[->, shorten <=-6pt] ($(SIMULATION.east) + (0mm, 0mm)$) to (PERF.west);
        \draw[->, rounded corners] (PERF.north) -- +(0mm, 3mm) -| node[above, very near start]{\scriptsize no} (SWEEP.north);
        \draw[->] (PERF.east) -- node[above]{\scriptsize yes} +(1, 0);
    \end{circuitikz}
    
    \caption{\footnotesize Conventional procedure of analog circuit design, which involves tremendous efforts to sweep in the parameter space to find the optimal design given design specifications. The design space contains possible parameters such as transistors, resistors, etc. Design specifications contain power consumption, bandwidth, etc.}
    \vspace{-4mm}
    \label{fig:procedure:conventional}
\end{figure}
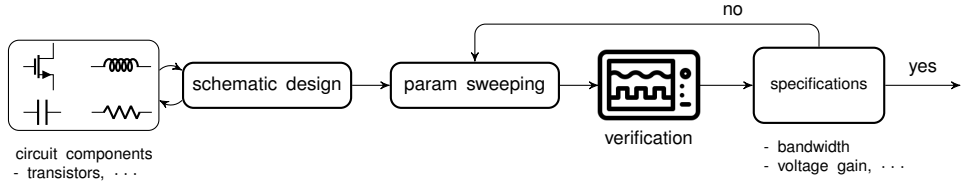

\textbf{Conventional analog circuit design} typically requires considerable effort and human involvement when searching the design space, including circuit topology and each circuit's parameters. 
Given a set of design specifications (e.g., power consumption, bandwidth, etc.), designers usually need first to decide the circuit topology and then conduct \emph{circuit sizing} (i.e., parameter sweeping on each individual component), as shown in Figure \ref{fig:procedure:conventional}. The process is extremely time-consuming when the circuit comprises a large number of components. Among all the phases within the design of analog circuits, the complete schematic-level design is the most time-consuming part which is currently handled by human experts in industry and academia in a temporally inefficient manner. 

\textbf{A few prior works} investigate machine learning algorithms to automate analog circuit design \cite{DATE_AutoCkt,ICML_Analog,IEEE_Angel,MDPI_AnalogReview, DAC_GCNRL, DAC19_CombOpt, ICCAD19_BagNet}. 
In general, these methods train an ML model to learn the mapping between design specifications and actual circuit parameters. 
\cite{ICML_Analog} shows that a standard neural network trained with carefully filtered design parameters
can effectively predict the circuit parameters given design specifications. 
Methods such as AutoCkt \cite{DATE_AutoCkt} and L2DC \cite{NIPS18_L2DC} further adopt reinforcement learning framework to train neural networks for effective circuit sizing given design specifications. 
Furthermore, GNN-based methods \cite{DAC_GCNRL, ICML19_CircuitGNN} propose to use a graph neural network to better capture the circuit topology and improve the accuracy of circuit sizing. 
Methods such as AnGeL \cite{IEEE_Angel} try to use neural networks on a homogeneous combination of circuits with identical functions by breaking them down into stages of sub-circuits, thereby shrinking the design space and reducing the training complexity. 

Despite progress in the prior works, the current ML-driven design is still in its early stages.
In particular, current methods mainly show a proof of concept on homogeneous circuits and systems such as two-stage voltage amplifiers. It still remains unclear how to scale current NN-based solutions on real-world analog systems with different components. In particular, a real-world system usually consists of heterogeneous components that perform different functions, such as a transmitter with an oscillator followed by a power amplifier. Such systems introduce more complex mapping from inputs to outputs and exhibit high non-linearity. 
Furthermore, a high-quality dataset, that includes various complex circuits, is indispensable as a pivotal element in the ML-assisted circuit design. However, such a dataset is still not developed and accessible in the current ML-assisted circuit design field. 
Additionally, given the nascent state of ML-assisted circuit design, effective and universal metrics are highly anticipated to assess different ML-assisted solutions. 

\textbf{In this work}, we first introduce \dataset{}, a multi-level circuit dataset and benchmark for training ML algorithms to assist various types of analog circuit design. In particular, the dataset consists of seven pivotal analog and radio-frequency circuits: common-source voltage amplifier, two-stage voltage amplifier, cascode voltage amplifier, low-noise amplifier, power amplifier, voltage-controlled oscillator, and mixer. In addition, the dataset also consists of complex heterogeneous systems with multiple cascaded circuit components, which has not been investigated in prior works. 
\dataset{} is a comprehensive collection of circuit parameters and simulated performance metrics from an accurate commercial simulator, including advanced metrics not seen in any prior works. Hence, it can serve as a foundation for developing robust ML models for analog and radio-frequency circuit design. 
In addition to the dataset, we also conduct comprehensive evaluations on the benchmark dataset using various models. The models evaluated range from conventional machine learning algorithms, such as random forest, to modern neural networks. 
The evaluations show that on the benchmark dataset, ML algorithms have the potential to learn circuit design with relatively small errors compared to design specifications.

\section{Problem Statement}\label{sec:problem}

In this work, we investigate the capabilities of machine learning algorithms in automating analog and radio-frequency circuit design. In particular, we evaluate machine learning algorithms on \emph{homogeneous} and \emph{heterogeneous} circuits. 
We define homogeneous circuits that compromise multiple circuits with identical functions. For instance, a two-stage voltage amplifier can be seen as a homogeneous circuit with two cascaded single-stage voltage amplifiers (Fig \ref{fig:circuit:tsa}). 
On the other hand, we define heterogeneous circuits that compromise multiple circuit blocks with different functions, such as a transmitter with a voltage-controlled oscillator and a power amplifier (Fig \ref{fig:transmitter}).

\textbf{Machine Learning-Assisted Design.}
ML-assisted design usually adopts a reversed design flow compared to conventional circuit design. In particular, ML-assisted design takes a performance metric vector, $\bm{y}$, as inputs and uses a machine learning model $\mathcal{M}$ to predict a set of circuit parameters $\bm{x}$ as
\begin{equation}
    \bm{x} = \mathcal{M}(\bm{y}),
\end{equation}
The performance vector, $\bm{y}$, may contain DC power consumption, bandwidth, voltage gain, etc, which varies with circuit types. On the other hand, the circuit parameter vector, $\bm{x}$, describes the quantitative values of every component within the circuit, including resistances, capacitances, transistor widths, etc.
Compared to conventional circuit design, ML-assisted methods directly learn the mapping from design specifications to circuit parameters, eliminating the need for parameter sweeping on $\bm{x}$ to find the solutions that meet the design specification in $\bm{y}$. Therefore, the whole design process can be significantly simplified. 

Various ML models such as multilayer perceptrons (MLPs), transformers, and conventional methods such as decision trees can be used to model the mapping from inputs to outputs. Simple circuits, such as a common-source voltage amplifier, can be modeled with small models. However, large circuit systems with high non-linearity require more complex models. 
As a key contribution in this work, we investigate the performance of different models on diverse circuits, ranging from a basic common-source amplifier to complex systems such as a transmitter with multiple heterogeneous circuit blocks. 

\section{Dataset}\label{sec:dataset}

In this section, we present our dataset collection procedure. 
\dataset{} consists of two types of circuits: homogeneous circuits and heterogeneous circuits. A homogeneous circuit only consists of one type of circuit, such as a common-source voltage amplifier. Heterongenous circuits may consist of two or more types of circuit blocks with different functions, such as a transmitter with a voltage-controlled oscillator (VCO) and a power amplifier (PA). 

For all circuits, we adopt the procedure as in Figure \ref{fig:dataset:proc} to generate data. For a circuit, we first design a schematic using Cadence tools \cite{Cadence02}. We defer details of circuit schematics in Appendix \ref{appx:circuit}. We then identify all key circuit parameters in the schematic that can affect the circuit performance. For each parameter, we set a value range, $[\texttt{beg}, \texttt{end}]$, and sweep the value with a small step size. 
For each parameter set, we run a Cadence simulator to obtain the simulation results and calculate performance metrics. Each parameter set and the corresponding performance will be saved as one row in the dataset. 
After sweeping all parameters, we obtain a dataset with all possible design points. As last, we split the dataset into train and test sets for ML model training and testing. 

\input{./fig/fig_dataset_proc}

\subsection{Datasets for Homogeneous Circuit Blocks}


We collect seven commonly used analog and radio-frequency circuits: common-source voltage amplifier (CSVA), cascode voltage amplifier (CVA), two-stage voltage amplifier (TSVA), low-noise amplifier (LNA), mixer, voltage-controlled oscillator (VCO), and power amplifier (PA). Schematics of these circuits are provided in Appendix \ref{appx:circuit}. These homogeneous circuits are the building units that are seen in real-world complex heterogeneous systems. 

Among these circuits, the analog voltage amplifiers, including the common-source voltage amplifier (CSVA), cascode voltage amplifier (CVA), and two-stage voltage amplifier (TSVA), are essential for voltage amplification, a critical function in most analog circuits and feedback systems \cite{razavi_design_2017}. CSVA is a versatile and widely used component in analog and radio-frequency circuit design. It receives input at the \emph{gate} terminal and generates amplified output at the \emph{drain} terminal (See Figure \ref{fig:circuit:csva}). 
By combining the \emph{common-source} (CS) and \emph{common-gate} (CG) stages, CVA can provide enhanced gain and improved bandwidth over CSVA, which is suitable for high-frequency applications (Figure \ref{fig:circuit:cascode}). 
TSVA further improves the output swings in the cascode configuration and obtains high gains (Figure \ref{fig:circuit:tsa}). 

In addition to voltage amplifiers, the dataset also covers other types of circuits in radio-frequency applications, including the low-noise amplifier (LNA), mixer, voltage-controlled oscillator (VCO), and power amplifier (PA). 
In particular, the cascode LNA in a radio-frequency receiver front-end provides substantial power gain while maintaining low noise across a wide bandwidth range (Figure \ref{fig:circuit:lna}). 
An active mixer is used for frequency modulation with conversion gain in radio-frequency transmitters and receivers. 
VCO generates a periodic signal with frequency tuned across a wide range controlled by a voltage signal. Owing to the low phase noise and power consumption, the cross-coupled VCO has become a prevalent configuration to provide sustainable oscillation (Figure \ref{fig:circuit:vco}). 
The two-stage differential PA, the most power-intensive building block in the radio-frequency transmitter, plays a crucial role in delivering significant power to the transmitting antenna without compromising efficiency (Figure \ref{fig:circuit:pa}).

For each circuit, we select several circuit parameters that can greatly affect the design performance, as listed in Table \ref{tab:circuitparam:basic}. Based on the complexity of each circuit, a different number of parameters are considered in the simulation. The channel length of each transistor is fixed at 45 nm to simplify the design space and mitigate short-channel effects. 
These parameters are also the targets that an ML algorithm needs to predict given design specifications. 

\begin{figure}[!htb]
    \centering
    \begin{subfigure}{0.45\textwidth}
        \centering
            \hspace{-0.2\textwidth}
            \scalebox{0.6}{    \begin{circuitikz}[line width=0.2mm]
        \draw (0,0) node[oscillator,anchor=south](VCO){}; 
        \draw (VCO.south) node[below=5pt] {\parbox{5cm}{\centering $\text{Voltage-Controlled}$ \\ \vspace{2pt} $\text{Oscillator}$}};

        \draw (VCO.east) -- ([xshift=0.3cm]VCO.east);
        \draw ([xshift=0.3cm]VCO.east) to[amp,name=buffer] ++(3.2,0);
        \draw (buffer.west) node[inputarrow,scale=1.5]{};
        \draw (buffer.south) node[below=5pt]{$\text{Buffer}$};
        \draw ([xshift=-0.1cm,yshift=-0.2cm]buffer.center) node[rectangle,draw,densely dashed,minimum width=7.9cm, minimum height=4.35cm](box){};
        \draw (box.center) node[below=46pt] {$\texttt{28\,GHz Signal Generator - Amplifier}$};

        \draw (buffer.east) to[amp,name=PA1] ++(3.5,0);
        \draw (PA1.west) node[inputarrow,scale=1.5]{};
        \draw (PA1.east) to[amp,name=PA2,xshift=-1.5cm,fill=white] ++(2.5,0);
        \draw (PA2.south) node[below=5pt]{$\text{Power Amplifier}$};

        \draw ([xshift=0.15cm]PA2.east) node[txantenna,anchor=center](Txantenna){};
        \draw ([xshift=0.5cm]Txantenna.north) node[above=10pt] {$\text{RF Signal}$};
    \end{circuitikz}}
        \caption{Transmitter system}
        \label{fig:transmitter}
    \end{subfigure}
    \hspace{20pt}
    \begin{subfigure}{0.45\textwidth}
        \centering
            \hspace{-0.2\textwidth}
            \scalebox{0.6}{ \begin{circuitikz}[line width=0.2mm]
        \draw (0,0) node[rxantenna,xscale=-1](rxantenna){};
        \draw ([xshift=-0.6cm]rxantenna.north) node[above=10pt]{\text{RF Signal}};

        \draw (rxantenna.center) to[amp,name=LNA] ++(1,0);
        \draw (LNA.west) node[inputarrow,scale=1.5]{};
        \draw (LNA.south) node[below=5pt] {\parbox{2cm}{\centering $\text{Low-Noise}$ \\ \vspace{2pt} $\text{Amplifier}$}};
    
        \draw ([xshift=1.1cm]LNA.east) node[mixer,anchor=west](Mixer){}; 
        \draw (LNA.east) -- (Mixer.west);
        \draw (Mixer.west) node[inputarrow, scale=1.5]{};
        \draw (Mixer.north) node[inputarrow,rotate=-90,scale=1.5]{};
        \draw (Mixer.south) node[below=5pt]{$\text{Mixer}$};
        \draw ([xshift=0.8cm,,yshift=-0.2cm]Mixer.center) node[rectangle,draw,densely dashed,minimum width=8cm, minimum height=4.35cm](box){};
        \draw (box.center) node[below=46pt] {$\texttt{28\,GHz Frequency Conversion Chain}$};

        \draw (Mixer.north) to[short,-*]($(Mixer.north)+(0,0.4)$) node[above=1pt] {\parbox{2cm}{\centering $\text{LO}$ \\ \vspace{1pt} $\texttt{(27.9\,GHz)}$}};
        
        \draw ([xshift=1.4cm]Mixer.east) node[lowpassshape](LPF){};
        \draw (LPF.west) node[inputarrow, scale=1.5]{};
        \draw (LPF.south) node[below=5pt]{\parbox{2cm}{\centering $\text{Low-Pass}$ \\ \vspace{2pt} $\text{Filter}$}};
        \draw (Mixer.east) -- (LPF.west);

        \draw ([xshift=1cm]LPF.east) to[amp,name=IFAmp] ++(1,0);
        \draw (IFAmp.west) node[inputarrow, scale=1.5]{};
        \draw (IFAmp.south) node[below=5pt]{\parbox{2cm}{\centering $\text{Cascode}$ \\ \vspace{2pt} $\text{Amplifier}$}};
        \draw (LPF.east) -- (IFAmp.west);
        \draw (IFAmp.east) to[short, -o] ($(IFAmp.east)+(0.8,0)$) node[right=5pt] {$\text{IF Signal}$};
    \end{circuitikz}}
        \caption{Receiver system}
        \label{fig:receiver}
    \end{subfigure}
    \caption{28\,GHz wireless transceiver circuits. (a) Transmitter architecture involving VCO and PA. Buffer used here to sustain system stability; (b) Receiver architecture comprising LNA, Mixer, and CVA. Low-Pass Filter deployed here to filter out the undesired high frequency components.}
    \label{fig:transceiver}
\end{figure}
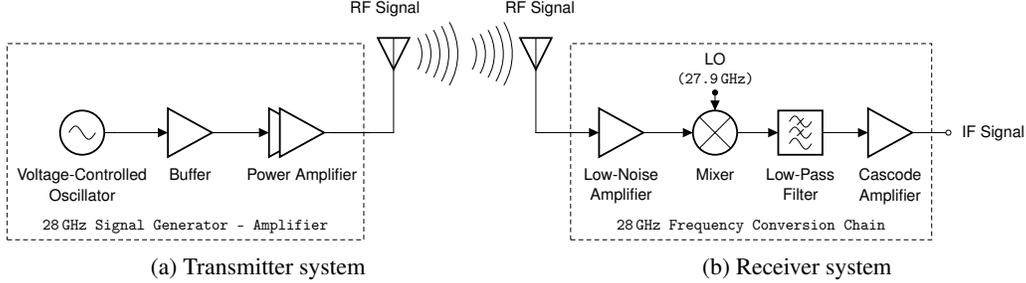

\subsection{Datasets for Complex Heterogeneous Systems}

In addition to basic homogeneous circuits, the work further investigates complex real-world millimeter-wave (mm-wave) circuit systems that contain multiple circuit blocks with different functions, as illustrated in Figure \ref{fig:transceiver} \cite{IEEE_mmwave}.
In particular, we investigate a transmitter and a receiver operating at 28\,GHz, which are commonly used in high-speed communication systems for sending and receiving mm-wave signals \cite{Millimiter_wave_circuit}.

For a transmitter, we combine the voltage-controlled oscillator (VCO) and power amplifier (PA) as a typical \emph{signal generator – amplifier} system (Figure \ref{fig:transmitter}) \cite{razavi_RFIC}. The system first generates a periodic signal via a VCO with tunable frequency, and then amplifies the signal by the PA with substantial power gain. 
For a receiver, we establish a classical \emph{frequency conversion} chain 
by integrating the low-noise amplifier (LNA) with a mixer and cascode voltage amplifier (CVA) (Figure \ref{fig:receiver}) \cite{li2002multi}. With a signal received from an antenna, an LNA is first applied to amplify the weak input signal without introducing undesired noise. 
Then, a mixer is involved in converting the signal from radio frequency to intermediate frequency (IF) \cite{razavi_receiver}. 
The output IF signal is then amplified by a CVA that serves as an IF amplifier for further processing. 
There are two additional blocks, buffer and low-pass filter, shown in the transmitter and receiver. As their topology and parameters are usually fixed, we do not optimize them in the pipeline.
Table \ref{tab:circuitparam:complex} lists design parameters to be optimized and the performance metrics to be examined. 

Compared to basic homogeneous circuits, these heterogeneous circuits comprise a large parameter space to be optimized. Moreover, these systems exhibit increased non-linearity and intricate trade-offs between each block, leading to further challenges in learning the mapping from performance metrics to design parameters.

\section{Evaluations}\label{sec:eval}

With the dataset collected from diverse circuits and complex radio-frequency systems, in this section, we train and evaluate multiple ML algorithms on the dataset and investigate their strengths and weaknesses.

\textbf{Models}. We test five different models: multi-layer perceptrons (MLPs), Transformers, support vector regression (SVRs), random forest (RF), and K-nearest neighbors (KNNs). For all models, we feed the model with performance metrics and let the model predict the design parameters. In particular, MLPs/Transformers/SVRs act as a mapping function from input metrics to output design parameters, while KNNs locate the design parameters with performance close to input metrics. 
In addition, RF regressors \cite{ML01_RandomForest} combine the predictions from multiple decision trees and output the mean of their predictions to create a more accurate and stable prediction.
The transformer model \cite{NIPS23_Attention} with the implementation based on \cite{arXiv19_Bert}, consists of one embedding layer, several encoder layers, and one fully connected layer for predicting a vector of circuit parameters. 
The multi-layer perception (MLP) model consists of seven fully connected layers, each intermediate layer having a rectified linear unit (ReLU) activation function. 
For support vector regressor (SVR), considering that SVR is a single-output regressor, we create multiple SVRs to predict all the circuit parameters. In particular, we fit one SVR per target parameter. To enhance non-linearity, we adopt the $\texttt{rbf}$ kernel \cite{MIT04_Kernel} for each SVR model.
Details of each model are provided in Appendix \ref{appx:model}.

\textbf{Metrics}. In model training, we use $\ell_1$ loss as the objective function that measures the distance between the predicted parameters and the desired parameters. In model evaluation, we further run a Cadence simulator given the predicted parameters and obtain the performance, $\hat{\bm{y}}$. We calculate the relative error compared to the desired performance specified in the dataset, $\bm{y}$. We report an individual error on each metric as
\begin{equation}\label{eq:relerror}
    i\text{th metric}: err_i = \| \bm{y}_i - \hat{\bm{y}}_i \| / \bm{y}_i
\end{equation}

\begin{figure}[!htb]
\centering
\begin{tikzpicture}
\node[draw=none, fill=gray!10, minimum height=3cm, text width=3.7cm, rounded corners, label=above:{\footnotesize train}, inner sep=1mm, align=left](train){\tiny
\begin{lstlisting}[language=Python]
def Train(model, trainset):
  for _ in range(maxIter):
    x, y = loadData(trainset)
    xPred = model(y)
    loss = calcLoss(x, xPred)
    # update model
    model.fit(x, y)
\end{lstlisting}
};

\draw[-, dashed, purple, opacity=.3] ($(train.north east) + (2mm, 0mm)$) -- ($(train.south east) + (2mm, 0mm)$);

\node[right=.4cm of train, draw=none, fill=gray!10, minimum height=3cm, text width=4.7cm, rounded corners, label=above:{\footnotesize evaluation}, inner sep=1mm](test){\tiny
\begin{lstlisting}[language=Python]
def Eval(model, evalset, simulator):
  for _ in range(maxIter):
    x, y = loadData(evalset)
    xPred = model(y)
    # circuit workflow
    # simulator: simulator class with all circuit information
    yPred = simulator.run(xPred)
    error = calcError(y, yPred)
\end{lstlisting}
};

\draw[-, dashed, purple, opacity=.3] ($(test.north east) + (2mm, 0mm)$) -- ($(test.south east) + (2mm, 0mm)$);

\node[right=.4cm of test, draw=none, fill=gray!10, minimum height=3cm, text width=4cm, rounded corners, label=above:{\footnotesize simulation}, inner sep=1mm](simulator){\tiny
\begin{lstlisting}[language=Python]
class Simulator():
  ...
  ...
  def run(self, xPred):
    alterCircParam(xPred)
    # call Cadence simulator
    call(cadenceCommand)
    yPred = parseResults()
\end{lstlisting}
};
\end{tikzpicture}
\caption{An end-to-end model training and evaluation pipeline.}
\label{fig:pipeline}
\end{figure}
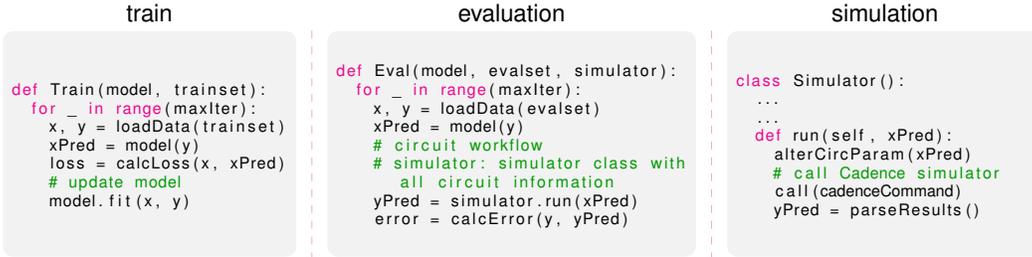

\textbf{End-to-End Training and Evaluation}. 
Our codebase provides an end-to-end model training and evaluation pipeline, as shown in Figure \ref{fig:pipeline}. It enables a smooth interaction between the ML workflow and the analog circuit workflow. During the training stage, we simply follow the standard ML workflow to load data and train the model. During the evaluation phase, we first obtain the predicted parameters via the ML workflow and then call Cadence simulator to compute the actual performance and the relative error compared to the desired value. 
Importantly, by including a Cadence simulator in the evaluation pipeline, we can accurately obtain metrics with inherent randomness based on advanced analyses, such as the noise figure in a low-noise amplifier and the phase noise in a voltage-controlled oscillator, which were not seen in prior works. More details are provided in Table \ref{tab:circuitparam:basic} and \ref{tab:circuitparam:complex}.

As the performance metrics and parameters targeted by the models have different ranges, we first apply the preprocessing and normalize the data to $[-1, 1]$. 
For training data splitting, we randomly sample 90 percent of the data points as the training dataset, and 10 percent as the test dataset. 
We train the neural networks (transformer and MLP) for 100 epochs using the Adam optimizer \cite{arXiv17_Adam} with a learning rate of 0.001. Each training is conducted several times to ensure that our methods are robust to random seeds.

\subsection{Homogeneous Circuit Blocks}\label{sec:eval:single}
We first evaluate ML algorithms on homogeneous circuits. In particular, we show the results of MLP, Transformer, SVR, and RF on a two-stage voltage amplifier (TSVA), Mixer, voltage-controlled oscillator (VCO), and a power amplifier (PA) in the main paper, and defer other results of other circuits in Appendix \ref{appx:additional}.

To better present the results, we plot the distribution of relative errors (See Eq(\ref{eq:relerror})) of all metrics. We can directly evaluate different ML algorithms by comparing their error distributions in the same plot.
We'll begin by presenting the results of each individual circuit, followed by a discussion of some common observations across all circuits.

\textbf{Two-Stage Voltage Amplifier (TSVA)}. We can observe that all four algorithms can give a design with small errors on the DC power consumption. The reason is that the mapping from circuit parameters to power consumption is easy to predict. However for other metrics, some models fail. For instance, SVR models always generate circuits with large errors on the bandwidth. 
For the voltage gain, circuits predicted by all four models have large relative errors compared to the desired specifications. 

\pgfplotstableread[col sep=comma]{VoltageGainTransformer.csv}{\loadedtableA}
\pgfplotstableread[col sep=comma]{BandwidthTransformer.csv}{\loadedtableB}
\pgfplotstableread[col sep=comma]{PowerConsumptionTransformer.csv}{\loadedtableC}

\pgfplotstableread[col sep=comma]{VoltageGainMLP.csv}{\loadedtableAA}
\pgfplotstableread[col sep=comma]{BandwidthMLP.csv}{\loadedtableBB}
\pgfplotstableread[col sep=comma]{PowerConsumptionMLP.csv}{\loadedtableCC}

\pgfplotstableread[col sep=comma]{VoltageGainSVR.csv}{\loadedtableAAA}
\pgfplotstableread[col sep=comma]{BandwidthSVR.csv}{\loadedtableBBB}
\pgfplotstableread[col sep=comma]{PowerConsumptionSVR.csv}{\loadedtableCCC}

\pgfplotstableread[col sep=comma]{VoltageGainRF.csv}{\loadedtableAAAA}
\pgfplotstableread[col sep=comma]{BandwidthRF.csv}{\loadedtableBBBB}
\pgfplotstableread[col sep=comma]{PowerConsumptionRF.csv}{\loadedtableCCCC}

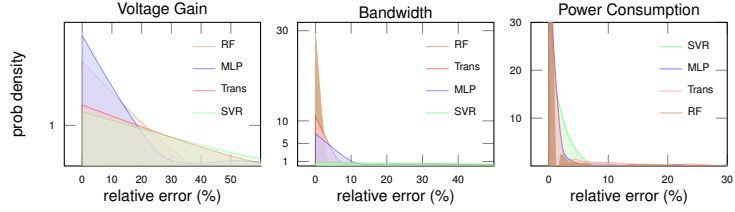
\begin{figure}[!htb]
    \centering
    \begin{tikzpicture}
        \begin{axis} [
            log origin=infty,
            title={\tiny Voltage Gain}, title style={yshift=-1.5ex},
            ticklabel style={font=\fontsize{4}{5}\selectfont},
            width=0.3\linewidth, height=.25\linewidth,
            xlabel={\tiny relative error (\%)}, x label style={at={(axis description cs:0.5,-0.1)},anchor=north},
            xmax=0.6, xtick={ 0, 0.1, 0.2, 0.3, 0.4, 0.5 }, xticklabels = {0, 10, 20, 30, 40, 50},
            ylabel={\tiny prob density}, ymin=0, y label style={at={(axis description cs:-0.15,.5)},anchor=south},
            ytick={ 1, 5 }, yticklabels={ 1, 5 },
            legend style={at={ (1.0,.6)}, anchor=east, draw=none, fill=none, font=\fontsize{4}{5}\selectfont },
            legend cell align={left}
        ]
            \addplot[thin, fill=brown!20, draw=brown!70, smooth, opacity=0.6] table [x=bin, y=cnt, col sep=comma] {\loadedtableAAAA} \closedcycle;
            \addlegendentry{RF}

            \addplot[thin, fill=blue!20, draw=blue!70, smooth, opacity=0.6] table [x=bin, y=cnt, col sep=comma] {\loadedtableAA} \closedcycle;
            \addlegendentry{MLP}
            
            \addplot[thin, fill=red!20, draw=red!70, smooth, opacity=0.6] table [x=bin, y=cnt, col sep=comma] {\loadedtableA} \closedcycle;
            \addlegendentry{Trans}

            \addplot[thin, fill=green!20, draw=green!70, smooth, opacity=0.4] table [x=bin, y=cnt, col sep=comma] {\loadedtableAAA} \closedcycle;
            \addlegendentry{SVR}

        \end{axis}
    \end{tikzpicture}   
    \begin{tikzpicture}
        \begin{axis} [
            log origin=infty,
            title={\tiny Bandwidth}, title style={yshift=-1.5ex},
            ticklabel style={font=\fontsize{4}{5}\selectfont},
            width=0.3\linewidth, height=.25\linewidth,
            xlabel={\tiny relative error (\%)}, x label style={at={(axis description cs:0.5,-0.1)},anchor=north},
            xtick={ 0, 0.1, 0.2, 0.3, 0.4 }, xticklabels = {0, 10, 20, 30, 40}, scaled x ticks=false,
            xmax=0.5, ymin=0,
            ytick={ 1, 5, 10, 30 }, yticklabels={ 1, 5, 10, 30 },
            legend style={at={ (1.0,.6)}, anchor=east, draw=none, fill=none, font=\fontsize{4}{5}\selectfont },
            legend cell align={left}
        ]    

            \addplot[thin, fill=brown!20, brown=blue!70, smooth, opacity=0.6] table [x=bin, y=cnt, col sep=comma] {\loadedtableBBBB} \closedcycle;
            \addlegendentry{RF}

            \addplot[thin, fill=red!20, draw=red!70, smooth, opacity=0.7] table [x=bin, y=cnt, col sep=comma] {\loadedtableB} \closedcycle;
            \addlegendentry{Trans}
            
            \addplot[thin, fill=blue!20, draw=blue!70, smooth, opacity=0.6] table [x=bin, y=cnt, col sep=comma] {\loadedtableBB} \closedcycle;
            \addlegendentry{MLP}

            \addplot[thin, fill=green!20, draw=green!70, smooth, opacity=0.6] table [x=bin, y=cnt, col sep=comma] {\loadedtableBBB} \closedcycle;
            \addlegendentry{SVR}

        \end{axis}
    \end{tikzpicture} 
    \begin{tikzpicture}
        \begin{axis} [
            log origin=infty,
            title={\tiny Power Consumption}, title style={yshift=-1.5ex},
            ticklabel style={font=\fontsize{4}{5}\selectfont},
            width=0.3\linewidth, height=.25\linewidth,
            xlabel={\tiny relative error (\%)},  x label style={at={(axis description cs:0.5,-0.1)},anchor=north},
            xtick={ 0, 0.1, 0.2, 0.3}, xticklabels = {0, 10, 20, 30}, scaled x ticks=false,
            xmax=0.3, ymin=0, ymax=30,
            ytick={ 10, 20, 30 }, yticklabels={ 10, 20, 30 },
            legend style={at={ (1.0,.6)}, anchor=east, draw=none, fill=none, font=\fontsize{4}{5}\selectfont },
            legend cell align={left}
        ]      

            \addplot[thin, fill=green!20, draw=green!70, smooth, opacity=0.6] table [x=bin, y=cnt, col sep=comma] {\loadedtableCCC} \closedcycle;
            \addlegendentry{SVR}
            
            \addplot[thin, fill=blue!20, draw=blue!70, smooth, opacity=0.7] table [x=bin, y=cnt, col sep=comma] {\loadedtableCC} \closedcycle;
            \addlegendentry{MLP}

            \addplot[thin, fill=red!20, draw=red!50, smooth, opacity=0.6] table [x=bin, y=cnt, col sep=comma] {\loadedtableC} \closedcycle;
            \addlegendentry{Trans}

            \addplot[thin, fill=brown!20, brown=blue!70, smooth, opacity=0.6] table [x=bin, y=cnt, col sep=comma] {\loadedtableCCCC} \closedcycle;
            \addlegendentry{RF}

        \end{axis}
    \end{tikzpicture} 
    \caption{Two-Stage Voltage Amplifier}
    \label{fig:error:compare:twostage}
    \vspace{-3mm}
\end{figure}

\textbf{Mixer}. First we observe similar performance on the DC power consumption. All models generate circuits with very small errors on power consumption. However, advanced models such as MLPs and Transformers perform better on complex metrics. In particular, for the conversion gain, circuits predicted by MLPs and transformers result in $\leq 10\%$ error, which is significantly lower than SVR and RF. 

\pgfplotstableread[col sep=comma]{PowerConsumptionTransformer.csv}{\loadedtableA}
\pgfplotstableread[col sep=comma]{VoltageSwingTransformer.csv}{\loadedtableB}
\pgfplotstableread[col sep=comma]{ConversionGainTransformer.csv}{\loadedtableC}
\pgfplotstableread[col sep=comma]{NoiseFigureTransformer.csv}{\loadedtableD}

\pgfplotstableread[col sep=comma]{PowerConsumptionMLP.csv}{\loadedtableAA}
\pgfplotstableread[col sep=comma]{VoltageSwingMLP.csv}{\loadedtableBB}
\pgfplotstableread[col sep=comma]{ConversionGainMLP.csv}{\loadedtableCC}
\pgfplotstableread[col sep=comma]{NoiseFigureMLP.csv}{\loadedtableDD}

\pgfplotstableread[col sep=comma]{PowerConsumptionSVR.csv}{\loadedtableAAA}
\pgfplotstableread[col sep=comma]{VoltageSwingSVR.csv}{\loadedtableBBB}
\pgfplotstableread[col sep=comma]{ConversionGainSVR.csv}{\loadedtableCCC}
\pgfplotstableread[col sep=comma]{NoiseFigureSVR.csv}{\loadedtableDDD}

\pgfplotstableread[col sep=comma]{PowerConsumptionRF.csv}{\loadedtableAAAA}
\pgfplotstableread[col sep=comma]{VoltageSwingRF.csv}{\loadedtableBBBB}
\pgfplotstableread[col sep=comma]{ConversionGainRF.csv}{\loadedtableCCCC}
\pgfplotstableread[col sep=comma]{NoiseFigureRF.csv}{\loadedtableDDDD}

\begin{figure}[!htb]
    \centering
    \begin{tikzpicture}
        \begin{semilogyaxis} [
            log origin=infty,
            title={\tiny Power Consumption}, title style={yshift=-1.5ex},
            ticklabel style={font=\fontsize{4}{5}\selectfont},
            width=0.3\linewidth, height=.25\linewidth,
            xlabel={\tiny relative error (\%)}, x label style={at={(axis description cs:0.5,-0.1)},anchor=north},
            xtick={ 0, .02, .04, .05 }, xticklabels = {0, 2, 4, 5}, scaled x ticks=false,
            xmax=0.06,
            ylabel={\tiny prob density}, ymin=0, y label style={at={(axis description cs:-0.15,.5)},anchor=south},
            ytick={ 1, 10, 50, 200 }, yticklabels={ 0, 10, 50, 200 },
            legend style={at={ (1,.6)}, anchor=east, draw=none, fill=none, font=\fontsize{4}{5}\selectfont },
            legend cell align={left}
        ]
            \addplot[thin, fill=green!20, draw=green!70, smooth, opacity=0.6] table [x=bin, y=cnt, col sep=comma] {\loadedtableAAA} \closedcycle;
            \addlegendentry{SVR}

            \addplot[thin, fill=brown!20, draw=brown!70, smooth, opacity=0.6] table [x=bin, y=cnt, col sep=comma] {\loadedtableAAAA} \closedcycle;
            \addlegendentry{RF}

            \addplot[thin, fill=blue!20, draw=blue!70, smooth, opacity=0.6] table [x=bin, y=cnt, col sep=comma] {\loadedtableAA} \closedcycle;
            \addlegendentry{MLP}
            
            \addplot[thin, fill=red!20, draw=red!70, smooth, opacity=0.6] table [x=bin, y=cnt, col sep=comma] {\loadedtableA} \closedcycle;
            \addlegendentry{Trans}

        \end{semilogyaxis}
    \end{tikzpicture}   
    \begin{tikzpicture}
        \begin{semilogyaxis} [
            log origin=infty,
            title={\tiny IF Voltage Swing}, title style={yshift=-1.5ex},
            ticklabel style={font=\fontsize{4}{5}\selectfont},
            width=0.3\linewidth, height=.25\linewidth,
            xlabel={\tiny relative error (\%)}, x label style={at={(axis description cs:0.5,-0.1)},anchor=north},
            xtick={ 0, .05, .1, .15 }, xticklabels = {0, 5, 10, 15}, scaled x ticks=false,
            xmax=0.2, ymin=0,
            ytick={ 1, 10, 50, 200 }, yticklabels={ 1, 10, 50, 200 },
            legend style={at={ (1.0,.6)}, anchor=east, draw=none, fill=none, font=\fontsize{4}{5}\selectfont },
            legend cell align={left}
        ]    

            \addplot[thin, fill=brown!20, brown=blue!70, smooth, opacity=0.6] table [x=bin, y=cnt, col sep=comma] {\loadedtableBBBB} \closedcycle;
            \addlegendentry{RF}

            \addplot[thin, fill=green!20, draw=green!70, smooth, opacity=0.6] table [x=bin, y=cnt, col sep=comma] {\loadedtableBBB} \closedcycle;
            \addlegendentry{SVR}
            
            \addplot[thin, fill=blue!20, draw=blue!70, smooth, opacity=0.7] table [x=bin, y=cnt, col sep=comma] {\loadedtableBB} \closedcycle;
            \addlegendentry{MLP}
            
            \addplot[thin, fill=red!20, draw=red!70, smooth, opacity=0.8] table [x=bin, y=cnt, col sep=comma] {\loadedtableB} \closedcycle;
            \addlegendentry{Trans}

        \end{semilogyaxis}
    \end{tikzpicture} 
    \begin{tikzpicture}
        \begin{semilogyaxis} [
            log origin=infty,
            title={\tiny Conversion Gain}, title style={yshift=-1.5ex},
            ticklabel style={font=\fontsize{4}{5}\selectfont},
            width=0.3\linewidth, height=.25\linewidth,
            xlabel={\tiny relative error (\%)},  x label style={at={(axis description cs:0.5,-0.1)},anchor=north},
            xtick={ 0, 0.1, 0.2, 0.3}, xticklabels = {0, 10, 20, 30}, scaled x ticks=false,
            xmax=0.5, ymin=0,
            ytick={ 1, 10, 50, 200 }, yticklabels={ 1, 10, 50, 200 },
            legend style={at={ (1.0,.6)}, anchor=east, draw=none, fill=none, font=\fontsize{4}{5}\selectfont },
            legend cell align={left}
        ]      

            \addplot[thin, fill=brown!20, brown=blue!70, smooth, opacity=0.6] table [x=bin, y=cnt, col sep=comma] {\loadedtableCCCC} \closedcycle;
            \addlegendentry{RF}

            \addplot[thin, fill=green!20, draw=green!70, smooth, opacity=0.6] table [x=bin, y=cnt, col sep=comma] {\loadedtableCCC} \closedcycle;
            \addlegendentry{SVR}
            
            \addplot[thin, fill=blue!20, draw=blue!70, smooth, opacity=0.7] table [x=bin, y=cnt, col sep=comma] {\loadedtableCC} \closedcycle;
            \addlegendentry{MLP}
            
            \addplot[thin, fill=red!20, draw=red!50, smooth, opacity=0.8] table [x=bin, y=cnt, col sep=comma] {\loadedtableC} \closedcycle;
            \addlegendentry{Trans}

        \end{semilogyaxis}
    \end{tikzpicture} 
    \begin{tikzpicture}
        \begin{semilogyaxis} [
            log origin=infty,
            title={\tiny Noise Figure}, title style={yshift=-1.5ex},
            ticklabel style={font=\fontsize{4}{5}\selectfont},
            width=0.3\linewidth, height=.25\linewidth,
            xlabel={\tiny relative error (\%)}, x label style={at={(axis description cs:0.5,-0.1)},anchor=north},
            xtick={ 0, 0.05, 0.1, 0.15 }, xticklabels = {0, 5, 10, 15}, scaled x ticks=false,
            xmax=0.10, ymin=0,
            ytick={ 1, 10, 50, 200 }, yticklabels={ 1, 10, 50, 200 },
            legend style={at={ (1.0,.6)}, anchor=east, draw=none, fill=none, font=\fontsize{4}{5}\selectfont },
            legend cell align={left}
        ]            
            \addplot[thin, fill=green!20, draw=green!70, smooth, opacity=0.6] table [x=bin, y=cnt, col sep=comma] {\loadedtableDDD} \closedcycle;
            \addlegendentry{SVR}
            
            \addplot[thin, fill=brown!20, draw=brown!70, smooth, opacity=0.5] table [x=bin, y=cnt, col sep=comma] {\loadedtableDDDD} \closedcycle;
            \addlegendentry{RF}
            
            \addplot[thin, fill=blue!20, draw=blue!70, smooth, opacity=0.6] table [x=bin, y=cnt, col sep=comma] {\loadedtableDD} \closedcycle;
            \addlegendentry{MLP}
            
            \addplot[thin, fill=red!20, draw=red!70, smooth, opacity=0.6] table [x=bin, y=cnt, col sep=comma] {\loadedtableD} \closedcycle;
            \addlegendentry{Trans}
        \end{semilogyaxis}
    \end{tikzpicture} 
    \caption{Mixer}
    \label{fig:error:compare:Mixer}
    \vspace{-3mm}
\end{figure}

\textbf{Voltage-Controlled Oscillator (VCO)}. We observe that all four models generate circuits with small errors on all metrics except phase noise. Even MLPs, that perform well on other circuits, result in a long error tail on the phase noise. 
The reason is that the mapping from circuit parameters to phase noise is highly non-linear, which makes it difficult for all models to learn the relationship.

\pgfplotstableread[col sep=comma]{OscillationFrequencyTransformer.csv}{\loadedtableA}
\pgfplotstableread[col sep=comma]{OutputPowerTransformer.csv}{\loadedtableB}
\pgfplotstableread[col sep=comma]{PhaseNoiseTransformer.csv}{\loadedtableC}
\pgfplotstableread[col sep=comma]{TuningRangeTransformer.csv}{\loadedtableD}
\pgfplotstableread[col sep=comma]{PowerConsumptionTransformer.csv}{\loadedtableE}

\pgfplotstableread[col sep=comma]{OscillationFrequencyMLP.csv}{\loadedtableAA}
\pgfplotstableread[col sep=comma]{OutputPowerMLP.csv}{\loadedtableBB}
\pgfplotstableread[col sep=comma]{PhaseNoiseMLP.csv}{\loadedtableCC}
\pgfplotstableread[col sep=comma]{TuningRangeMLP.csv}{\loadedtableDD}
\pgfplotstableread[col sep=comma]{PowerConsumptionMLP.csv}{\loadedtableEE}

\pgfplotstableread[col sep=comma]{OscillationFrequencyRF.csv}{\loadedtableAAA}
\pgfplotstableread[col sep=comma]{OutputPowerRF.csv}{\loadedtableBBB}
\pgfplotstableread[col sep=comma]{PhaseNoiseRF.csv}{\loadedtableCCC}
\pgfplotstableread[col sep=comma]{TuningRangeRF.csv}{\loadedtableDDD}
\pgfplotstableread[col sep=comma]{PowerConsumptionRF.csv}{\loadedtableEEE}

\pgfplotstableread[col sep=comma]{OscillationFrequencySVR.csv}{\loadedtableAAAA}
\pgfplotstableread[col sep=comma]{OutputPowerSVR.csv}{\loadedtableBBBB}
\pgfplotstableread[col sep=comma]{PhaseNoiseSVR.csv}{\loadedtableCCCC}
\pgfplotstableread[col sep=comma]{TuningRangeSVR.csv}{\loadedtableDDDD}
\pgfplotstableread[col sep=comma]{PowerConsumptionSVR.csv}{\loadedtableEEEE}

\begin{figure}[!htb]
    \centering
    \begin{tikzpicture}
        \begin{semilogyaxis} [
            log origin=infty,
            title={\tiny Oscillation Frequency}, title style={yshift=-1.5ex},
            ticklabel style={font=\fontsize{4}{5}\selectfont},
            width=0.3\linewidth, height=.25\linewidth,
            xlabel={\tiny relative error (\%)}, x label style={at={(axis description cs:0.5,-0.1)},anchor=north},
            xtick={ 0, 0.1, 0.2, 0.3, 0.4, 0.5 }, xticklabels = {0, 10, 20, 30, 40, 50},
            xmax=0.20,
            ylabel={\tiny prob density}, ymin=0, y label style={at={(axis description cs:-0.15,.5)},anchor=south},
            ytick={ 1, 10, 50, 200 }, yticklabels={ 0, 10, 50, 200 },
            legend style={at={ (1.0,.6)}, anchor=east, draw=none, fill=none, font=\fontsize{4}{5}\selectfont },
            legend cell align={left}
        ]
            \addplot[thin, fill=brown!20, draw=brown!70, smooth, opacity=0.6] table [x=bin, y=cnt, col sep=comma] {\loadedtableAAA} \closedcycle;
            \addlegendentry{RF}

            \addplot[thin, fill=blue!20, draw=blue!70, smooth, opacity=0.6] table [x=bin, y=cnt, col sep=comma] {\loadedtableAA} \closedcycle;
            \addlegendentry{MLP}
            
            \addplot[thin, fill=red!20, draw=red!70, smooth, opacity=0.8] table [x=bin, y=cnt, col sep=comma] {\loadedtableA} \closedcycle;
            \addlegendentry{Trans}

            \addplot[thin, fill=green!20, draw=green!70, smooth, opacity=0.6] table [x=bin, y=cnt, col sep=comma] {\loadedtableAAAA} \closedcycle;
            \addlegendentry{SVR}

        \end{semilogyaxis}
    \end{tikzpicture}   
    \begin{tikzpicture}
        \begin{semilogyaxis} [
            log origin=infty,
            title = {\tiny Output Power}, title style={yshift=-1.5ex},
            ticklabel style={font=\fontsize{4}{5}\selectfont},
            width=0.3\linewidth, height=.25\linewidth,
            xlabel={\tiny relative error (\%)}, x label style={at={(axis description cs:0.5,-0.1)},anchor=north},
            xtick={ 0, 0.02 }, xticklabels = {0, 2}, scaled x ticks=false,
            xmax=0.025, ymin=0,
            ytick={ 1, 10, 50, 200 }, yticklabels={ 1, 10, 50, 200 },
            legend style={at={ (1.0,.6)}, anchor=east, draw=none, fill=none, font=\fontsize{4}{5}\selectfont },
            legend cell align={left}
        ]    

            \addplot[thin, fill=green!20, draw=green!70, smooth, opacity=0.6] table [x=bin, y=cnt, col sep=comma] {\loadedtableBBBB} \closedcycle;
            \addlegendentry{SVR}

            \addplot[thin, fill=brown!20, draw=brown!70, smooth, opacity=0.8] table [x=bin, y=cnt, col sep=comma] {\loadedtableBBB} \closedcycle;
            \addlegendentry{RF}
            
            \addplot[thin, fill=blue!20, draw=blue!70, smooth, opacity=0.6] table [x=bin, y=cnt, col sep=comma] {\loadedtableBB} \closedcycle;
            \addlegendentry{MLP}
            
            \addplot[thin, fill=red!20, draw=red!70, smooth, opacity=0.6] table [x=bin, y=cnt, col sep=comma] {\loadedtableB} \closedcycle;
            \addlegendentry{Trans}

        \end{semilogyaxis}
    \end{tikzpicture} 
    \begin{tikzpicture}
        \begin{semilogyaxis} [
            log origin=infty,
            title={\tiny Phase Noise}, title style={yshift=-1.5ex},
            ticklabel style={font=\fontsize{4}{5}\selectfont},
            width=0.3\linewidth, height=.25\linewidth,
            xlabel={\tiny relative error (\%)},  x label style={at={(axis description cs:0.5,-0.1)},anchor=north},
            xtick={ 0, 0.1, 0.2, 0.3}, xticklabels = {0, 10, 20, 30}, scaled x ticks=false,
            xmax=0.4, ymin=0,
            ytick={ 1, 10, 50, 200 }, yticklabels={ 1, 10, 50, 200 },
            legend style={at={ (1.0,.6)}, anchor=east, draw=none, fill=none, font=\fontsize{4}{5}\selectfont },
            legend cell align={left}
        ]      

            \addplot[thin, fill=green!20, draw=green!70, smooth, opacity=0.6] table [x=bin, y=cnt, col sep=comma] {\loadedtableCCCC} \closedcycle;
            \addlegendentry{SVR}
            
            \addplot[thin, fill=blue!20, draw=blue!70, smooth, opacity=0.7] table [x=bin, y=cnt, col sep=comma] {\loadedtableCC} \closedcycle;
            \addlegendentry{MLP}
            
            \addplot[thin, fill=red!20, draw=red!50, smooth, opacity=0.8] table [x=bin, y=cnt, col sep=comma] {\loadedtableC} \closedcycle;
            \addlegendentry{Trans}

            \addplot[thin, fill=brown!20, draw=brown!70, smooth, opacity=0.6] table [x=bin, y=cnt, col sep=comma] {\loadedtableCCC} \closedcycle;
            \addlegendentry{RF}

        \end{semilogyaxis}
    \end{tikzpicture} 
    \begin{tikzpicture}
        \begin{axis} [
            log origin=infty,
            title={\tiny Tuning Range}, title style={yshift=-1.5ex},
            ticklabel style={font=\fontsize{4}{5}\selectfont},
            width=0.3\linewidth, height=.25\linewidth,
            xlabel={\tiny relative error (\%)}, x label style={at={(axis description cs:0.5,-0.1)},anchor=north},
            xtick={ 0, 0.05, 0.1, 0.15 }, xticklabels = {0, 5, 10, 15}, scaled x ticks=false,
            xmax=0.25, ymin=0.01,
            ytick={ 1, 10, 50, 200 }, yticklabels={ 1, 10, 50, 200 },
            legend style={at={ (1.0,.6)}, anchor=east, draw=none, fill=none, font=\fontsize{4}{5}\selectfont },
            legend cell align={left}
        ] 
            
            \addplot[thin, fill=blue!20, draw=blue!70, smooth, opacity=0.4] table [x=bin, y=cnt, col sep=comma] {\loadedtableDD} \closedcycle;
            \addlegendentry{MLP}
            
            \addplot[thin, fill=red!20, draw=red!70, smooth, opacity=0.4] table [x=bin, y=cnt, col sep=comma] {\loadedtableD} \closedcycle;
            \addlegendentry{Trans}

            \addplot[thin, fill=green!20, draw=green!70, smooth, opacity=0.6] table [x=bin, y=cnt, col sep=comma] {\loadedtableDDDD} \closedcycle;
            \addlegendentry{SVR}

            \addplot[thin, fill=brown!20, draw=brown!70, smooth, opacity=0.6] table [x=bin, y=cnt, col sep=comma] {\loadedtableDDD} \closedcycle;
            \addlegendentry{RF}
        \end{axis}
    \end{tikzpicture} 
    \caption{Voltage-Controlled Oscillator}
    \label{fig:error:compare:VCO}
\end{figure}
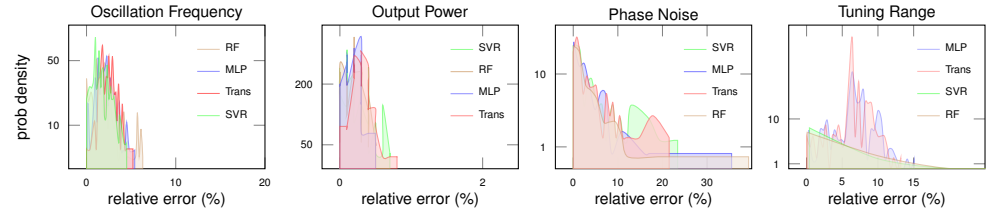

\textbf{Power Amplifier (PA)}. A power amplifier is one of the most complex circuits, a complexity that is reflected in the obtained results. Exception for the DC power consumption, all other metrics observe long error tails for all four models. In particular, the circuits predicted by SVR diverge significantly on the metrics of $\text{S}_{\text{11}}$ and $\text{S}_{\text{22}}$. Therefore, we did not include them in the figure. 

\input{./plot/plot_error_pa}

\textbf{Summary}. Across all circuits examined in this section, we observe that ML models perform well in learning the relationship between circuit parameters and simple performance metrics such as the DC power consumption. Importantly, the observation holds regardless of the complexity of the circuits. The reason is that the simple performance metrics usually exhibit a linear relation with circuit parameters, which makes it easy to predict. 
However, as the relationship becomes more non-linear, learning the relationships tends to be more challenging, even for MLPs and Transformers. For instance, voltage gain in a two-stage voltage amplifier, conversion gain in a mixer, phase noise in VCO, and performance metrics in PA are highly non-linear in relation to the circuit parameters. As a result, circuits predicted by ML models do not result in performance very close to the desired specifications. 

\subsection{Heterogeneous Radio-Frequency Systems}\label{sec:eval:complex}

Moving from homogenous circuits to heterogenous ones, the number of performance metrics and circuit parameters increases. Furthermore, as heterogeneous circuits comprise multiple circuits, the relationship between circuit parameters and performance is further complicated. 
Therefore, evaluations on heterogeneous circuits provide more insights into how ML algorithms learn to predict complex circuits. 

\textbf{Transmitter}. In a transmitter system comprising VCO and PA, we can observe that as the circuit becomes more complex, circuit parameters are harder to predict to meet the desired specification. Models such as SVR and RF lack sufficient capacity to predict complex systems, resulting in large errors. MLPs and transformers, on the other hand, predict circuits with much smaller errors. However, they still struggle to give a design that meets certain complex specifications, such as output power. 

For the metrics of individual components, circuits predicted by MLPs and transformers still result in smaller errors compared to SVRs and RF. One interesting observation is that, as more performance metrics are involved in learning a heterogeneous circuit, the trained models can generate circuits with small errors on individual metrics (e.g., the tuning range of VCO compared to Fig \ref{fig:error:compare:VCO}).  

\input{./plot/plot_error_transmitter}

\textbf{Receiver}. We further evaluate ML models on a receiver system comprising an LNA, a mixer, and a cascode voltage amplifier. Compared to the transmitter system, circuits in the receiver are less complex. Therefore, models such as MLPs and transmitters can predict circuits with smaller errors compared to the desired specifications. 
Besides, since the training dataset of the receiver includes more circuit parameters than the transmitter, the trained models exhibit better generalization performance with more training data points. 

\pgfplotstableread[col sep=comma]{PowerConsumptionMLP.csv}{\tableA}
\pgfplotstableread[col sep=comma]{VoltageGainMLP.csv}{\tableB}
\pgfplotstableread[col sep=comma]{NoiseFigureMLP.csv}{\tableC}

\pgfplotstableread[col sep=comma]{PowerConsumptionTransformer.csv}{\tableAA}
\pgfplotstableread[col sep=comma]{VoltageGainTransformer.csv}{\tableBB}
\pgfplotstableread[col sep=comma]{NoiseFigureTransformer.csv}{\tableCC}

\pgfplotstableread[col sep=comma]{PowerConsumptionSVR.csv}{\tableAAA}
\pgfplotstableread[col sep=comma]{VoltageGainSVR.csv}{\tableBBB}
\pgfplotstableread[col sep=comma]{NoiseFigureSVR.csv}{\tableCCC}

\pgfplotstableread[col sep=comma]{PowerConsumptionRF.csv}{\tableAAAA}
\pgfplotstableread[col sep=comma]{VoltageGainRF.csv}{\tableBBBB}
\pgfplotstableread[col sep=comma]{NoiseFigureRF.csv}{\tableCCCC}

\begin{figure}[!htb]
    \centering
    \begin{tikzpicture}
        \begin{semilogyaxis} [
            log origin=infty,
            title={\tiny Power Consumption}, title style={yshift=-1.5ex},
            ticklabel style={font=\fontsize{4}{5}\selectfont},
            width=0.3\linewidth, height=.25\linewidth,
            xlabel={\tiny relative error (\%)}, x label style={at={(axis description cs:0.5,-0.1)},anchor=north},
            xtick={ 0, .02, .04 }, xticklabels = {0, 2, 4}, scaled x ticks=false,
            xmax=0.05,
            ylabel={\tiny prob density}, ymin=0, y label style={at={(axis description cs:-0.15,.5)},anchor=south},
            ytick={ 1, 10, 50, 200 }, yticklabels={ 0, 10, 50, 200 },
            legend style={at={ (1,.6)}, anchor=east, draw=none, fill=none, font=\fontsize{4}{5}\selectfont },
            legend cell align={left}
        ]

            \addplot[thin, fill=green!40, draw=green!70, smooth, opacity=0.6] table [x=bin, y=cnt, col sep=comma] {\tableAAA} \closedcycle;
            \addlegendentry{SVR}

            \addplot[thin, fill=blue!20, draw=blue!70, smooth, opacity=0.6] table [x=bin, y=cnt, col sep=comma] {\tableA} \closedcycle;
            \addlegendentry{MLP}

            \addplot[thin, fill=red!20, draw=red!70, smooth, opacity=0.6] table [x=bin, y=cnt, col sep=comma] {\tableAA} \closedcycle;
            \addlegendentry{Trans}

            \addplot[thin, fill=brown!20, draw=brown!70, smooth, opacity=0.7] table [x=bin, y=cnt, col sep=comma] {\tableAAAA} \closedcycle;
            \addlegendentry{RF}

        \end{semilogyaxis}
    \end{tikzpicture}   
    \begin{tikzpicture}
        \begin{semilogyaxis} [
            log origin=infty,
            title={\tiny Voltage Gain}, title style={yshift=-1.5ex},
            ticklabel style={font=\fontsize{4}{5}\selectfont},
            width=0.3\linewidth, height=.25\linewidth,
            xlabel={\tiny relative error (\%)}, x label style={at={(axis description cs:0.5,-0.1)},anchor=north},
            xtick={ 0, .1, .2 }, xticklabels = {0, 10, 20}, scaled x ticks=false,
            xmax=0.3, ymin=0,
            ytick={ 1, 10, 50, 200 }, yticklabels={ 1, 10, 50, 200 },
            legend style={at={ (1.0,.6)}, anchor=east, draw=none, fill=none, font=\fontsize{4}{5}\selectfont },
            legend cell align={left}
        ]    

            \addplot[thin, fill=green!20, draw=green!70, smooth, opacity=0.7] table [x=bin, y=cnt, col sep=comma] {\tableBBB} \closedcycle;
            \addlegendentry{SVR}

            \addplot[thin, fill=brown!20, brown=blue!70, smooth, opacity=0.6] table [x=bin, y=cnt, col sep=comma] {\tableBBBB} \closedcycle;
            \addlegendentry{RF}

            \addplot[thin, fill=blue!20, draw=blue!70, smooth, opacity=0.6] table [x=bin, y=cnt, col sep=comma] {\tableB} \closedcycle;
            \addlegendentry{MLP}

            \addplot[thin, fill=red!20, draw=red!70, smooth, opacity=0.6] table [x=bin, y=cnt, col sep=comma] {\tableBB} \closedcycle;
            \addlegendentry{Trans}

        \end{semilogyaxis}
    \end{tikzpicture} 
    \begin{tikzpicture}
        \begin{semilogyaxis} [
            log origin=infty,
            title={\tiny Noise Figure}, title style={yshift=-1.5ex},
            ticklabel style={font=\fontsize{4}{5}\selectfont},
            width=0.3\linewidth, height=.25\linewidth,
            xlabel={\tiny relative error (\%)},  x label style={at={(axis description cs:0.5,-0.1)},anchor=north},
            xtick={ 0, .02, .04, .06}, xticklabels = {0, 2, 4, 6}, scaled x ticks=false,
            xmax=0.1, ymin=0,
            ytick={ 1, 10, 50, 200 }, yticklabels={ 1, 10, 50, 200 },
            legend style={at={ (1.0,.6)}, anchor=east, draw=none, fill=none, font=\fontsize{4}{5}\selectfont },
            legend cell align={left}
        ]      

            \addplot[thin, fill=green!20, draw=green!70, smooth, opacity=0.6] table [x=bin, y=cnt, col sep=comma] {\tableCCC} \closedcycle;
            \addlegendentry{NF: SVR}

            \addplot[thin, fill=brown!20, brown=blue!70, smooth, opacity=0.6] table [x=bin, y=cnt, col sep=comma] {\tableCCCC} \closedcycle;
            \addlegendentry{RF}

            \addplot[thin, fill=blue!20, draw=blue!70, smooth, opacity=0.7] table [x=bin, y=cnt, col sep=comma] {\tableC} \closedcycle;
            \addlegendentry{MLP}
            
            \addplot[thin, fill=red!20, draw=red!50, smooth, opacity=0.8] table [x=bin, y=cnt, col sep=comma] {\tableCC} \closedcycle;
            \addlegendentry{Trans}

        \end{semilogyaxis}
    \end{tikzpicture} 
    \caption{Receiver (LNA, Mixer, and Cascode)}
    \label{fig:error:compare:receiver}
\end{figure}
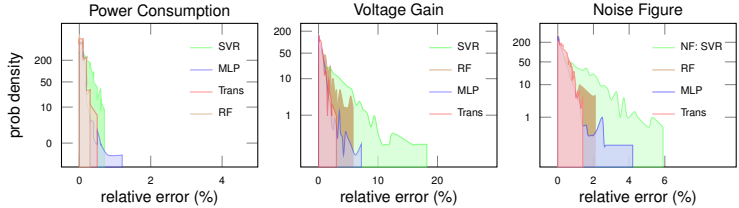

\textbf{Summary}. Throughout the evaluations on a transmitter and a receiver system, we can observe how the performance of ML algorithms is affected by the complexity of circuits. From a receiver to a transmitter, the relationship between circuit parameters and the performance becomes more non-linear. As a result, it is more challenging to predict a circuit that meets design specifications. 


\subsection{Learning from Homogeneous to Heterogeneous Circuits}

Some prior works argue that models trained on small circuits can be extended to large complex circuits \cite{IEEE_Angel}. However, we want to point out that it does not apply to heterogeneous circuits. 
As we revealed in Figure \ref{fig:error:transmitter:individual}, training an ML model on an individual circuit that is a part of a complex system results in a much worse learning result. 
In particular, by training an MLP model with PA-related metrics, we simulate the transmitter system and only calculate the error distribution on PA's metrics. We observe that the predicted circuits behave differently as specified in the datasets, resulting in large errors compared to the desired values.
The rationale behind this is that a PA within a transmitter system operates at a different point than when it operates independently, due to the load effects caused by other circuits in the system. Therefore, a model that simply learns a circuit alone cannot generalize well on a large system. 



\pgfplotstableread[col sep=comma]{LargeSignalPowerGainPAInd.csv}{\loadedtableLargeSignalPowerGainPAInd}
\pgfplotstableread[col sep=comma]{DrainEfficiencyPAInd.csv}{\loadedtableDrainEfficiencyPAInd}
\pgfplotstableread[col sep=comma]{PAEPAInd.csv}{\loadedtablePAEPAInd}

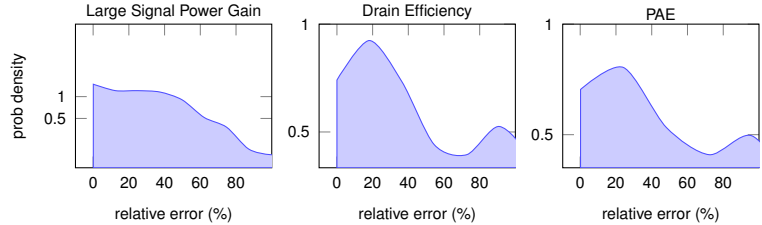
\begin{figure}[!htb]
    \centering
    \begin{tikzpicture}
        \begin{semilogyaxis} [
            log origin=infty,
            title={\tiny Large Signal Power Gain}, title style={yshift=-1.5ex},
            ticklabel style={font=\tiny},
            width=0.3\linewidth, height=.25\linewidth,
            xlabel={\tiny relative error (\%)},
            xtick={ 0, 0.2, 0.4, 0.6, 0.8 }, xticklabels = {0, 20, 40, 60, 80},
            xmax=1.0, ymax= 10,
            ylabel={\tiny prob density}, ymin=0,
            ytick={.5, 1}, yticklabels={ 0.5, 1},
            legend style={at={ (1.1,1.1)}, anchor=east, draw=none, fill=none },
            legend cell align={left}
        ]

            \addplot[thin, fill=blue!20, draw=blue!70, smooth] table [x=bin, y=cnt, col sep=comma] {\loadedtableLargeSignalPowerGainPAInd} \closedcycle;

        \end{semilogyaxis}
    \end{tikzpicture} 
    \begin{tikzpicture}
        \begin{semilogyaxis} [
            log origin=infty,
            title={\tiny Drain Efficiency}, title style={yshift=-1.5ex},
            ticklabel style={font=\tiny},
            width=0.3\linewidth, height=.25\linewidth,
            xlabel={\tiny relative error (\%)},
            xtick={ 0, 0.2, 0.4, 0.6, 0.8 }, xticklabels = {0, 20, 40, 60, 80},
            xmax=1.0, ymax= 1,
            ytick={ 0.5, 1 }, yticklabels={ 0.5, 1 },
            legend style={at={ (1.1,1.1)}, anchor=east, draw=none, fill=none },
            legend cell align={left}
        ]
            \addplot[thin, fill=blue!20, draw=blue!70, smooth] table [x=bin, y=cnt, col sep=comma] {\loadedtableDrainEfficiencyPAInd} \closedcycle;

        \end{semilogyaxis}
    \end{tikzpicture}   
    \begin{tikzpicture}
        \begin{semilogyaxis} [
            log origin=infty,
            title={\tiny PAE}, title style={yshift=-1.5ex},
            ticklabel style={font=\tiny},
            width=0.3\linewidth, height=.25\linewidth,
            xlabel={\tiny relative error (\%)},
            xtick={ 0, 0.2, 0.4, 0.6, 0.8 }, xticklabels = {0, 20, 40, 60, 80},
            xmax=1.0, ymax= 1,
            ytick={ 0.5, 1 }, yticklabels={ 0.5, 1 },
            legend style={at={ (.9,1.1)}, anchor=east, draw=none, fill=none },
            legend cell align={left}
        ]
            \addplot[thin, fill=blue!20, draw=blue!70, smooth] table [x=bin, y=cnt, col sep=comma] {\loadedtablePAEPAInd} \closedcycle;

        \end{semilogyaxis}
    \end{tikzpicture} 

    \caption{Performance of MLP on PA in a transmitter using only PA-related metrics. The accuracy is significantly affected by the limited number of metrics and the effects of circuits in the transmitter.}
    \label{fig:error:transmitter:individual}
    \vspace{-3mm}
\end{figure}



\section{Conclusion}\label{sec:conclusion}

In this work, we propose a multi-level benchmark dataset for analog and radio-frequency circuit design. The proposed dataset, \dataset{}, covers homogeneous and heterogeneous circuits. Homogeneous circuits comprise one or multiple circuits with identical functions, while heterogeneous circuits comprise circuits with different functions. 
We evaluate various machine learning algorithms on the benchmark datasets, including multi-layer perceptrons (MLPs), transformers, and support vector regression (SVRs).
The evaluations provide a comprehensive overview of the strengths and weaknesses of each method. In a word, MLPs and Transformers usually give better designs compared to other methods, especially for complex circuits. 
We also reveal that for complex circuits, further optimization of model design and training are still needed to improve the design. 

Heterogeneous circuits investigated in the paper only cover typical radio-frequency transmitter and receiver systems. However, we will consider extending our research to other complex systems.


{
	\bibliographystyle{plain}
	\bibliography{./references}
}

\newpage 

\appendix

\section{Circuits}\label{appx:circuit}
In this section, we provide details on the examined homogeneous and heterogeneous circuits in Section \ref{sec:dataset} and \ref{sec:eval}. As mentioned in the body of the paper, time-consuming parametric sweeps are necessary for designing analog and radio frequency circuits. Often, a small subset of design parameters can satisfy the thresholds for design metrics. From a circuit design perspective, the limited number of parameter combinations that satisfy the metrics happens due to: 1) complexity of transistor models and the sensitivity of their operation to surrounding circuit elements, circuit configuration, bias conditions, size of the transistor, etc. 2) inherent trade-offs among various design metrics in each circuit. In what follows, for each circuit, in addition to the schematics and set of design parameters and metrics, multiple important design trade-offs are listed within colored boxes adjacent to the schematics. The same color is reserved for a trade-off if it is present in more than one circuit. 
\begin{figure}[!htb]
    \centering
    \begin{minipage}[b]{0.45\textwidth}
     \centering
      \begin{circuitikz}[scale=0.9, transform shape]
            \ctikzset{tripoles/mos style/arrows}
            \ctikzset{capacitors/scale=0.4}
            \ctikzset{resistors/scale=0.6}
            \ctikzset{transistors/scale=0.9}
            \ctikzset{grounds/scale=0.8}
            
            \draw (0,0) node [nmos, anchor = G](M1) {$\text{W}_\text{N}$};
            \draw (M1.G) to[short, -o] ($(M1.G)-(0.3,0)$) node[left] {$\text{V}_\text{gate}$};
            \draw ([yshift=0.1cm]M1.D) to[short, -o] ($(M1.D)+(1.5,0.1)$) node[right] {$\text{V}_\text{out}$};
            \draw ([xshift=1.1cm,yshift=0.1cm]M1.D) -- ([xshift=1.1cm,yshift=-0.2cm]M1.D);
            \draw ([xshift=1.1cm,yshift=-0.2cm]M1.D) to[C,name=C] ++(0,-1);
            \draw ([yshift=-0.43cm]C.east) node[ground](GND){};
            
            \draw (M1.D) to[R,l_=$\text{R}_\text{D}$,name=R] ++(0,1.8);
            
            \draw [line width=0.5mm] ([yshift=0.9cm, xshift=-1.7cm]R.north) -- ([yshift=0.9cm, xshift=1.7cm]R.north) node[above, near end] {$\textbf{V}_\textbf{DD}$};
            
            \draw (M1.S) node[ground](GND){};
            \end{circuitikz}
    \end{minipage}
    \hspace{-5mm}
    \raisebox{0.2cm}{
    \begin{minipage}[b]{0.35\textwidth}
        \centering
        \begin{tcolorbox}[colframe=black,colback=white,boxrule=0.5pt, width=0.8\textwidth,title=\scriptsize\centering\bfseries Homogeneous,
        fonttitle=\small\bfseries\color{black}, coltitle=black, colbacktitle=white]
            \scriptsize
            \begin{tcolorbox}[colframe=black,colback=red!30,boxrule=0pt,left=0pt,right=0pt,top=0pt,bottom=0pt]
            \centering 
                Gain-bandwidth tradeoff
            \end{tcolorbox}
            \begin{tcolorbox}[colframe=black,colback=green!30,boxrule=0pt,left=0pt,right=0pt,top=0pt,bottom=0pt]
            \centering
                Gain-swing tradeoff
            \end{tcolorbox}
            \begin{tcolorbox}[colframe=black,colback=blue!30,boxrule=0pt,left=0pt,right=0pt,top=0pt,bottom=0pt]
            \centering
                Gain-power tradeoff
            \end{tcolorbox}
        \end{tcolorbox}
    \end{minipage}
    }
    \vspace{10pt}
    \caption{Common-Source Voltage Amplifier.}
    \label{fig:circuit:csva}
\end{figure}
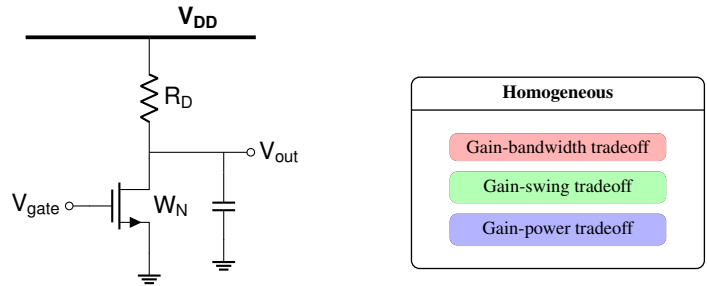
\begin{figure}[!htb]
    \centering
    \begin{minipage}[b]{0.45\textwidth}
     \centering
      \begin{circuitikz}[scale=0.85, transform shape]
        \ctikzset{tripoles/mos style/arrows}
        \ctikzset{capacitors/scale=0.4}
        \ctikzset{resistors/scale=0.6}
        \ctikzset{transistors/scale=0.8}
        \ctikzset{grounds/scale=0.8}
        
        \draw (0,0) node [nmos, anchor = G](M1){$\text{W}_\text{N1}$};
        \draw (M1.G) to[short, -o] ($(M1.G)-(0.3,0)$) node[left] {$\text{V}_\text{in}\text{+}$};
        
        \draw (5,0) node [nmos, anchor = G, xscale=-1](M3){\ctikzflipx{$\text{W}_\text{N1}$}};
        \draw (M3.G) to[short, -o] ($(M3.G)+(0.3,0)$) node[right] {$\text{V}_\text{in}\text{-}$};
        
        \draw (M1.D) -- ++(0,0.3) node[nmos, anchor=S, xscale=1](M2) {$\text{W}_\text{N2}$};
        \draw (M2.G) to[short, -*] ($(M2.G)-(0.3,0)$) node[left] {$\text{V}_\text{b1}$};
        \draw (M2.D) to[short, -o] ($(M2.D)+(1.3,0)$) node[above] {$\text{V}_\text{out}\text{+}$};
        \draw ([xshift=1cm]M2.D) to[C, name=C1] ++(0,-1.2);
        \draw ([yshift=-0.4cm]C1.east) node[ground](GND){};
        
        \draw (M3.D) -- ++(0,0.3) node[nmos, anchor=S, xscale=-1](M4) {\ctikzflipx{$\text{W}_\text{N2}$}};
        \draw (M4.G) to[short, -*] ($(M4.G)+(0.3,0)$) node[right] {$\text{V}_\text{b1}$};
        \draw (M4.D) to[short, -o] ($(M4.D)-(1.3,0)$) node[above] {$\text{V}_\text{out}\text{-}$};
        \draw ([xshift=-1cm]M4.D) to[C, name=C2] ++(0,-1.2);
        \draw ([yshift=-0.4cm]C2.east) node[ground](GND){};
        
        \draw (M2.D) to[R,l=$\text{R}_\text{D}$,name=R1] ++(0,1.6);
        \draw (M4.D) to[R,l_=$\text{R}_\text{D}$,name=R2] ++(0,1.6);
        
        \draw [line width=0.5mm] ([yshift=0.8cm, xshift=-1.6cm]R1.north) -- ([yshift=0.8cm,xshift=1.6cm]R2.north) node[above, near end] {$\textbf{V}_\textbf{DD}$};
        
        \draw (M1.S) -- ++(0,-0.2);
        \draw (M3.S) -- ++(0,-0.2);
        \draw ([yshift=-0.2cm]M1.S) -- ++(1.8,0) node [nmos, anchor = D](M5){$\text{W}_\text{N3}$};
        \draw (M5.G) to[short, -*] ($(M5.G)-(0.4,0)$) node[left] {$\text{V}_\text{b2}$};
        \draw ([yshift=-0.2cm]M3.S) -- (M5.D);
        
        \draw (M5.S) node[ground](GND){};
    \end{circuitikz}
    \end{minipage}
    \raisebox{1.5cm}{
    \begin{minipage}[b]{0.35\textwidth}
        \centering
        \begin{tcolorbox}[colframe=black,colback=white,boxrule=0.5pt, width=0.8\textwidth,title=\scriptsize\centering\bfseries Homogeneous,
        fonttitle=\small\bfseries\color{black}, coltitle=black, colbacktitle=white]
            \scriptsize
            \begin{tcolorbox}[colframe=black,colback=red!30,boxrule=0pt,left=0pt,right=0pt,top=0pt,bottom=0pt]
            \centering
                Gain-bandwidth tradeoff
            \end{tcolorbox}
            \begin{tcolorbox}[colframe=black,colback=green!30,boxrule=0pt,left=0pt,right=0pt,top=0pt,bottom=0pt]
            \centering
                Gain-swing tradeoff
            \end{tcolorbox}
            \begin{tcolorbox}[colframe=black,colback=blue!30,boxrule=0pt,left=0pt,right=0pt,top=0pt,bottom=0pt]
            \centering
                Gain-power tradeoff
            \end{tcolorbox}
        \end{tcolorbox}
    \end{minipage}
    }
    \vspace{10pt}
    \caption{Cascode Voltage Amplifier.}
    \label{fig:circuit:cascode}
\end{figure}
\begin{figure}[!htb]
    \centering
    \begin{minipage}[b]{0.5\textwidth}
     \centering
      \begin{circuitikz}[scale=0.75, transform shape]
        \ctikzset{tripoles/mos style/arrows}
        \ctikzset{tripoles/pmos style/nocircle}
        \ctikzset{capacitors/scale=0.4}
        \ctikzset{transistors/scale=0.8}
        \ctikzset{grounds/scale=0.8}

        \draw (0,0) node [nmos,anchor=G](M1){$\text{W}_\text{N1}$};
        \draw (M1.G) to[short,-o] ($(M1.G)-(0.05,0)$) node[left]{$\text{V}_\text{in}\text{+}$};

        \draw (3.5,0) node [nmos,anchor=G,xscale=-1](M5){\ctikzflipx {$\text{W}_\text{N1}$}};
        \draw (M5.G) to[short,-o]($(M5.G)+(0.05,0)$) node[right]{$\text{V}_\text{in}\text{-}$};

        \draw (M1.D) --++(0,0.3) node[nmos,anchor=S](M2){$\text{W}_\text{N1}$};
        \draw (M2.G) to[short,-*] ($(M2.G)-(0.05,0)$) node[left]{$\text{V}_\text{b1}$};

        \draw (M5.D) --++(0,0.3) node[nmos,anchor=S,xscale=-1](M6){\ctikzflipx{$\text{W}_\text{N1}$}};
        \draw (M6.G) to[short,-*] ($(M6.G)+(0.05,0)$) node[right]{$\text{V}_\text{b1}$};

        \draw (M2.D) --++(0,0.5) node [pmos,nocircle,arrowmos,anchor=D](M3){$\text{W}_\text{P1}$};
        \draw (M3.G) to[short,-*] ($(M3.G)-(0.05,0)$) node[left]{$\text{V}_\text{b2}$};

        \draw (M3.D) --++(-1.5,0) node [pmos,nocircle,arrowmos,anchor=G,xscale=-1](M9){\ctikzflipx{$\text{W}_\text{P2}$}};

        \draw ([xshift=0.3cm]M9.G) to[C, l={$\text{C}_\text{1}$},name=C1] ++(0,-0.8);
        \draw ([yshift=-0.35cm]C1.east) -- ([yshift=-0.19cm]M9.D);

        \draw (M9.D) --++(0,-0.5) node [nmos,anchor=D,xscale=-1](M10){\ctikzflipx{$\text{W}_\text{N2}$}};
        \draw (M10.G) to[short,-*] ($(M10.G)+(0.05,0)$) node[right]{$\text{V}_\text{b5}$};
        \draw (M10.D) to[short,-o] ($(M10.D)-(1.5,0)$) node[left] {$\text{V}_\text{out}\text{+}$};
        \draw ([xshift=-1.1cm]M10.D) to[C,name=C3] ++(0,-1);

        \draw (M6.D) --++(0,0.5) node [pmos,nocircle,arrowmos,anchor=D,xscale=-1](M7){\ctikzflipx{$\text{W}_\text{P1}$}};
        \draw (M7.G) to[short,-*] ($(M7.G)+(0.05,0)$) node[right]{$\text{V}_\text{b2}$};

        \draw (M7.D) --++(+1.5,0) node [pmos,nocircle,arrowmos,anchor=G](M11){$\text{W}_\text{P2}$};

        \draw ([xshift=-0.3cm]M11.G) to[C, l_={$\text{C}_\text{1}$},name=C2] ++(0,-0.8);
        \draw ([yshift=-0.35cm]C2.east) -- ([yshift=-0.19cm]M11.D);

        \draw (M11.D) --++(0,-0.5) node [nmos,anchor=D](M12){$\text{W}_\text{N2}$};
        \draw (M12.G) to[short,-*] ($(M12.G)-(0.05,0)$) node[left]{$\text{V}_\text{b5}$};
        \draw (M12.D) to[short,-o] ($(M12.D)+(1.5,0)$) node[right] {$\text{V}_\text{out}\text{-}$};
        \draw ([xshift=1.1cm]M12.D) to[C,name=C4] ++(0,-1);

        \draw (M3.S) --++(0,0.3) node[pmos,nocircle,arrowmos,anchor=D](M4){$\text{W}_\text{P1}$};
        \draw (M4.G) to[short,-*] ($(M4.G)-(0.05,0)$) node[left]{$\text{V}_\text{b3}$};

        \draw (M7.S) --++(0,0.3) node[pmos,nocircle,arrowmos,anchor=D,xscale=-1](M8){\ctikzflipx{$\text{W}_\text{P1}$}};
        \draw (M8.G) to[short,-*] ($(M8.G)+(0.05,0)$) node[right]{$\text{V}_\text{b3}$};

        \draw (M1.S) --++(0,-0.2);
        \draw (M5.S) --++(0,-0.2);
        \draw ([yshift=-0.2cm]M1.S) -- ++(1,0) node [nmos,anchor=D](M13){$\text{W}_\text{N3}$};
        \draw ([yshift=-0.2cm]M5.S) -- (M13.D);
        \draw (M13.G) to[short,-*] ($(M13.G)-(0.05,0)$) node[left]{$\text{V}_\text{b4}$};

        \draw (M4.S) -- ([yshift=0.4cm]M4.S);
        \draw (M8.S) -- ([yshift=0.4cm]M8.S);
        \draw (M9.S) -- ([yshift=2.55cm]M9.S);
        \draw (M11.S) -- ([yshift=2.55cm]M11.S);
        \draw [line width=0.5mm]([yshift=0.4cm, xshift=-3.5cm]M4.S)--([yshift=0.4cm, xshift=3.5cm]M8.S)
        node[above, near end]{$\textbf{V}_\textbf{DD}$};

        \draw ([yshift=-0.2cm]C3.east) node[ground](GND){};
        \draw ([yshift=-0.2cm]C4.east) node[ground](GND){};
        \draw (M10.S) node[ground](GND){};
        \draw (M12.S) node[ground](GND){};
        \draw (M13.S) node[ground](GND){};
    \end{circuitikz}
    \end{minipage}
    \hspace{10mm}
    \raisebox{2cm}{
    \begin{minipage}[b]{0.35\textwidth}
        \centering
        \begin{tcolorbox}[colframe=black,colback=white,boxrule=0.5pt, width=0.8\textwidth,title=\scriptsize\centering\bfseries Homogeneous,
        fonttitle=\small\bfseries\color{black}, coltitle=black, colbacktitle=white]
            \scriptsize
            \begin{tcolorbox}[colframe=black,colback=yellow!30,boxrule=0pt,left=0pt,right=0pt,top=0pt,bottom=0pt]
            \centering
                Noise-bandwidth tradeoff
            \end{tcolorbox}
            \begin{tcolorbox}[colframe=black,colback=red!30,boxrule=0pt,left=0pt,right=0pt,top=0pt,bottom=0pt]
            \centering
                Gain-bandwidth tradeoff
            \end{tcolorbox}
            \begin{tcolorbox}[colframe=black,colback=blue!30,boxrule=0pt,left=0pt,right=0pt,top=0pt,bottom=0pt]
            \centering
                Gain-power tradeoff
            \end{tcolorbox}
        \end{tcolorbox}
    \end{minipage}
    }
    \vspace{10pt}
    \caption{Two-Stage Voltage Amplifier.}
    \label{fig:circuit:tsa}
\end{figure}
\begin{figure}[!htb]
    \centering
    \begin{minipage}[b]{0.45\textwidth}
     \centering
      \begin{circuitikz}[scale=0.8, transform shape]
        \ctikzset{tripoles/mos style/arrows}
        \ctikzset{capacitors/scale=0.4}
        \ctikzset{resistors/scale=0.5}
        \ctikzset{inductors/scale=0.8, inductors/coils=4}
        \ctikzset{transistors/scale=0.8}
        \ctikzset{grounds/scale=0.8}

        \draw (0,0) node[nmos,anchor=G](M1){$\text{W}_\text{N1}$};

        \draw (M1.D)--++(0,0.3) node[nmos,anchor=S](M2){$\text{W}_\text{N2}$};
        \draw (M2.G)--++(-0.5,0);
        \draw ([xshift=-0.5cm]M2.G)--++(0,2.2);
        \draw (M2.D) --++(0.8,0);

        \draw (M2.D) to[L,l=$\text{L}_\text{d}$,name=Ld,mirror] ++(0,1.6);
        \draw (M1.G) to[L,l=$\text{L}_\text{g}$,name=Lg,mirror] ++(-1.4,0);
        \draw (M1.S) to[L,l=$\text{L}_\text{s}$,name=Ls] ++(0,-1.6);

        \draw ([xshift=0.8cm]M2.D) to[C,l_=$\text{C}_\text{1}$,name=C1] ++(1.6,0);
        \draw ([xshift=0.6cm]C1.east) to[C,name=C3] ++(0,-1.4);
        \draw ([xshift=0.75cm]C1.east) to[short,-o] ($(C1.east)+(1,0)$) node[right] {$\text{V}_\text{out}$};
        \draw ([xshift=-0.8cm]Lg.west) to[C,l=$\text{C}_\text{2}$,name=C2] ++(-1.4,0);
        \draw ([xshift=-0.75cm]C2.west) to[short,-o] ($(C2.west)-(0.9,0)$) node[left] {$\text{V}_\text{in}$};

        \draw ([xshift=0.8cm]M2.D) to[R,name=R3] ++(0,1.6);
        \draw ([xshift=-0.3cm]Lg.east) to[R,name=R2] ++(0,1.7);

        \draw ([xshift=-0.2cm,yshift=0.56cm]R2.east) node[nmos,anchor=G,xscale=-1](M3){};
        \draw (M3.D) to[R,name=R1] ++(0,1.45);
        \draw (M3.G) --++(0.2,0);
        \draw ([xshift=0.2cm]M3.G) --++(0,0.7);
        \draw ([xshift=0.2cm,yshift=0.7cm]M3.G) --++(-0.98,0);

        \draw [line width=0.5mm]([yshift=0.8cm,xshift=-4cm]Ld.north)--([yshift=0.8cm,xshift=2.5cm]Ld.north) node[above,near end]{$\textbf{V}_\textbf{DD}$};

        \draw (M3.S) node[ground](GND){};
        \draw ([yshift=-0.4cm]Ls.east) node[ground](GND){};
        \draw ([yshift=-0.6cm]C3.east) node[ground](GND){};
        
    \end{circuitikz}
    \end{minipage}
    \raisebox{1cm}{
    \begin{minipage}[b]{0.35\textwidth}
        \centering
        \begin{tcolorbox}[colframe=black,colback=white,boxrule=0.5pt, width=0.8\textwidth,title=\scriptsize\centering\bfseries Homogeneous,
        fonttitle=\small\bfseries\color{black}, coltitle=black, colbacktitle=white]
            \scriptsize
            \begin{tcolorbox}[colframe=black,colback=yellow!30,boxrule=0pt,left=0pt,right=0pt,top=0pt,bottom=0pt]
            \centering
                Noise-bandwidth tradeoff
            \end{tcolorbox}
            \begin{tcolorbox}[colframe=black,colback=red!30,boxrule=0pt,left=0pt,right=0pt,top=0pt,bottom=0pt]
            \centering
                Gain-bandwidth tradeoff
            \end{tcolorbox}
            \begin{tcolorbox}[colframe=black,colback=blue!30,boxrule=0pt,left=0pt,right=0pt,top=0pt,bottom=0pt]
            \centering
                Gain-power tradeoff
            \end{tcolorbox}
        \end{tcolorbox}
    \end{minipage}
    }
    \vspace{10pt}
    \caption{Low-Noise Amplifier.}
    \label{fig:circuit:lna}
\end{figure}
\begin{figure}[!htb]
    \centering  
    \begin{minipage}[b]{0.5\textwidth}
     \centering
       \begin{circuitikz}[scale=0.8, transform shape, american]
        \ctikzset{tripoles/mos style/arrows}
        \ctikzset{capacitors/scale=0.5}
        \ctikzset{resistors/scale=0.5}
        \ctikzset{inductors/scale=0.8, inductors/coils=4}
        \ctikzset{transistors/scale=0.8}
        \ctikzset{grounds/scale=0.8}
        \ctikzset{sources/scale=0.7}

        \draw (0,0) node[nmos,anchor=G](M1){$\text{W}_\text{N1}$};
        \draw (M1.G) to[short,-o] ($(M1.G)-(0.3,0)$) node[left] {$\text{V}_\text{RF}\text{+}$};

        \draw (5,0) node[nmos,anchor=G,xscale=-1](M2){\ctikzflipx{$\text{W}_\text{N1}$}};
        \draw (M2.G) to[short,-o] ($(M2.G)+(0.3,0)$) node[right] {$\text{V}_\text{RF}\text{-}$};

        \draw (M1.D) -- ([xshift=-0.8cm]M1.D);
        \draw ([xshift=-0.8cm]M1.D) --++(0,0.2) node[nmos,anchor=S](M3){$\text{W}_\text{N2}$};
        \draw (M3.G) to[short,-o] ($(M3.G)-(0.3,0)$) node[left] {$\text{V}_\text{LO}\text{+}$};
        \draw ([yshift=0.6cm]M3.D) to[short,-o] ($(M3.D)+(0.8,0.6)$) node[right] {$\text{V}_\text{IF}\text{+}$};

        \draw (M1.D) -- ([xshift=0.8cm]M1.D);
        \draw ([xshift=0.8cm]M1.D) --++(0,0.2) node[nmos,anchor=S,xscale=-1](M4){\ctikzflipx{$\text{W}_\text{N2}$}};
        \draw ([xshift=0.15cm]M4.G) to[short,-o] ($(M4.G)+(0.15,-0.6)$) node[below] {$\text{V}_\text{LO}\text{-}$};

        \draw (M2.D) -- ([xshift=-0.8cm]M2.D);
        \draw ([xshift=-0.8cm]M2.D) --++(0,0.2) node[nmos,anchor=S](M5){$\text{W}_\text{N2}$};
        \draw (M4.G) -- (M5.G);
        \draw (M3.D) -- (M5.D);

        \draw (M2.D) -- ([xshift=0.8cm]M2.D);
        \draw ([xshift=0.8cm]M2.D) --++(0,0.2) node[nmos,anchor=S,xscale=-1](M6){\ctikzflipx{$\text{W}_\text{N2}$}};
        \draw (M4.D) -- ([yshift=0.2cm]M4.D);
        \draw ([yshift=0.2cm]M4.D) -- ([yshift=0.2cm]M6.D);
        \draw (M6.G) to[short,-o] ($(M6.G)+(0.3,0)$) node[right] {$\text{V}_\text{LO}\text{+}$};
        \draw ([yshift=0.6cm]M6.D) to[short,-o] ($(M6.D)+(-0.8,0.6)$) node[left] {$\text{V}_\text{IF}\text{-}$};

        \draw (M3.D) -- ([yshift=0.3cm]M3.D);
        \draw ([yshift=0.3cm]M3.D) to[R,l_=$\text{R}$,name=R1] ++(0,1.5);
        \draw (M6.D) -- ([yshift=0.3cm]M6.D);
        \draw ([yshift=0.3cm]M6.D) to[R,l=$\text{R}$,name=R2] ++(0,1.5);

        \draw ([xshift=-0.7cm,yshift=-0.4cm]R1.west) to[C,l=$\text{C}$,name=C1] ++(0,1.45);
        \draw ([yshift=-0.65cm]C1.west) --([xshift=0.7cm,yshift=-0.65cm]C1.west);
        \draw ([xshift=0.7cm,yshift=-0.4cm]R2.west) to[C,l_=$\text{C}$,name=C2] ++(0,1.45);
        \draw ([yshift=-0.65cm]C2.west) --([xshift=-0.7cm,yshift=-0.65cm]C2.west);

        \draw [line width=0.5mm]([yshift=0.75cm,xshift=-1.8cm]R1.north)--([yshift=0.75cm,xshift=1.8cm]R2.north) node[above,near end]{$\textbf{V}_\textbf{DD}$};

        \draw (M1.S) -- ([yshift=-0.2cm]M1.S);
        \draw ([yshift=-0.2cm]M1.S) -- ([xshift=1.7cm, yshift=-0.2cm]M1.S);
        \draw (M2.S) -- ([yshift=-0.2cm]M2.S);
        \draw ([yshift=-0.2cm]M2.S) -- ([xshift=-1.75cm, yshift=-0.2cm]M2.S);
        \draw ([xshift=1.7cm, yshift=-0.2cm]M1.S) to[I,name=I1] ++(0,-1.4);

        \draw ([yshift=-0.2cm]I1.right) node[ground](GND){};

    \end{circuitikz}
    \end{minipage}
    \hspace{2mm}
    \raisebox{1cm}{
    \begin{minipage}[b]{0.35\textwidth}
        \centering
        \begin{tcolorbox}[colframe=black,colback=white,boxrule=0.5pt, width=0.8\textwidth,title=\scriptsize\centering\bfseries Homogeneous,
        fonttitle=\small\bfseries\color{black}, coltitle=black, colbacktitle=white]
            \scriptsize
            \begin{tcolorbox}[colframe=black,colback=yellow!30,boxrule=0pt,left=0pt,right=0pt,top=0pt,bottom=0pt]
            \centering
                Noise-bandwidth tradeoff
            \end{tcolorbox}
            \begin{tcolorbox}[colframe=black,colback=red!30,boxrule=0pt,left=0pt,right=0pt,top=0pt,bottom=0pt]
            \centering
                Gain-bandwidth tradeoff
            \end{tcolorbox}
            \begin{tcolorbox}[colframe=black,colback=blue!30,boxrule=0pt,left=0pt,right=0pt,top=0pt,bottom=0pt]
            \centering
                Gain-power tradeoff
            \end{tcolorbox}
        \end{tcolorbox}
    \end{minipage}
    }
    \vspace{10pt}
    \caption{Mixer.}
    \label{fig:circuit:mixer}
\end{figure}
\begin{figure}[!htb]
    \centering
    \begin{minipage}[b]{0.5\textwidth}
     \centering
       \begin{circuitikz}[scale=0.75, transform shape]
        \ctikzset{tripoles/mos style/arrows}
        \ctikzset{capacitors/scale=0.5}
        \ctikzset{resistors/scale=0.5}
        \ctikzset{inductors/scale=0.7, inductors/coils=4}
        \ctikzset{transistors/scale=0.8}
        \ctikzset{grounds/scale=0.8}
        \ctikzset{sources/scale=0.7}

        \draw (0,0) node[nmos,anchor=G,xscale=-1](M1){\ctikzflipx{$\text{W}_\text{N1}$}};
        \draw ([yshift=0.8cm]M1.D) to[short,-o] ($(M1.D)+(-1.5,0.8)$) node[left] {$\text{V}_\text{out}\text{+}$};
        
        \draw (1.5,0) node[nmos,anchor=G](M2){$\text{W}_\text{N1}$};
        \draw ([yshift=0.8cm]M2.D) to[short,-o] ($(M2.D)+(1.5,0.8)$) node[right] {$\text{V}_\text{out}\text{-}$};

        \draw ([xshift=-1.6cm,yshift=0.4cm]M1.D) node[nmos,anchor=G,rotate=-90](M3){};
        \draw ([xshift=-0.2cm]M3.S) node[anchor=south] {$\text{W}_\text{var}$};
        \draw (M3.G) -- ([xshift=1.6cm]M3.G);
        \draw (M3.D) -- ([yshift=-0.4cm]M3.D);
        \draw (M3.S) -- ([yshift=-0.4cm]M3.S);
        \draw ([yshift=-0.4cm]M3.D) -- ([yshift=-0.4cm]M3.S);

        \draw ([xshift=1.6cm,yshift=0.4cm]M2.D) node[nmos,anchor=G,rotate=-90](M4){};
        \draw ([xshift=0.2cm]M4.D) node[anchor=south] {$\text{W}_\text{var}$};
        \draw (M4.G) -- ([xshift=-1.6cm]M4.G);
        \draw (M4.D) -- ([yshift=-0.4cm]M4.D);
        \draw (M4.S) -- ([yshift=-0.4cm]M4.S);
        \draw ([yshift=-0.4cm]M4.D) -- ([yshift=-0.4cm]M4.S);
        \draw ([yshift=-1.2cm]M3.G) -- ([yshift=-2cm]M3.G);
        \draw ([yshift=-1.2cm]M4.G) -- ([yshift=-2cm]M4.G);
        \draw ([yshift=-2cm]M3.G) -- ([yshift=-2cm]M4.G);
        \draw ([xshift=1cm,yshift=-2cm]M3.G) to[short,-o] ($(M3.G)+(1,-2.4)$) node[below] {$\text{V}_\text{control}$};

        \draw (M1.S) -- (M2.S);
        \draw ([xshift=1.55cm]M1.S) -- ([xshift=1.55cm,yshift=-0.8cm]M1.S);
        \draw ([xshift=1.55cm,yshift=-0.8cm]M1.S) node[nmos,anchor=D](M5){$\text{W}_\text{N3}$};

        \draw ([xshift=-3.6cm]M5.G) node[nmos,anchor=G,xscale=-1](M6){\ctikzflipx{$\text{W}_\text{N2}$}};
        \draw (M6.G) -- (M5.G);

        \draw ([yshift=2cm]M6.D) to[american current source,name=I1,invert] ++(0,2);
        \draw ([yshift=0.7cm]I1.left) -- ([yshift=1.4cm]I1.left);
        \draw (M6.D) -- ([yshift=2cm]M6.D);
        \draw ([yshift=0.5cm]M6.D) -- ([xshift=1.5cm,yshift=0.5cm]M6.D);
        \draw ([xshift=1.5cm,yshift=0.5cm]M6.D) -- ([xshift=1.5cm,yshift=-0.6cm]M6.D);
        
        \draw (M1.G) -- ([xshift=0.2cm]M1.G);
        \draw (M2.G) -- ([xshift=-0.2cm]M2.G);
        \draw ([yshift=0.1cm]M1.D) -- ([xshift=0.8cm,yshift=0.1cm]M1.D);
        \draw ([yshift=0.1cm]M2.D) -- ([xshift=-0.8cm,yshift=0.1cm]M2.D);
        \draw ([xshift=0.2cm]M1.G) -- ([xshift=-0.8cm,yshift=0.1cm]M2.D);
        \draw ([xshift=-0.2cm]M2.G) -- ([xshift=0.8cm,yshift=0.1cm]M1.D);

        \draw (M1.D) -- ([yshift=1.2cm]M1.D);
        \draw ([yshift=1.2cm]M1.D) node[circ]{};
        \draw ([yshift=1.2cm]M1.D) to[R,l=$\text{R}_\text{p}$,name=R1] ++(0,1.5);
        \draw (M2.D) -- ([yshift=1.2cm]M2.D);
        \draw ([yshift=1.2cm]M2.D) node[circ]{};
        \draw ([yshift=1.2cm]M2.D) to[R,l=$\text{R}_\text{p}$,name=R2] ++(0,1.5);

        \draw ([xshift=-0.8cm,yshift=-0.75cm]R1.north) to[C,l=$\text{C}$,name=C1] ++(0,1.5);
        \draw ([yshift=-0.67cm]C1.west) -- ([xshift=0.9cm,yshift=-0.67cm]C1.west);
        \draw ([xshift=0.8cm,yshift=-0.75cm]R2.south) to[C,l=$\text{C}$,name=C2] ++(0,1.5);
        \draw ([yshift=-0.67cm]C2.west) -- ([xshift=-0.9cm,yshift=-0.67cm]C2.west);

        \draw ([xshift=0.8cm,yshift=-0.75cm]R1.south) to[L,l=$\text{L}$,name=L1] ++(0,1.5);
        \draw ([yshift=-0.45cm]L1.west) -- ([xshift=-0.9cm,yshift=-0.45cm]L1.west);
        \draw ([xshift=-0.8cm,yshift=-0.75cm]R2.north) to[L,l=$\text{L}$,name=L2] ++(0,1.5);
        \draw ([yshift=-0.45cm]L2.west) -- ([xshift=0.9cm,yshift=-0.45cm]L2.west);

        \draw [line width=0.5mm]([yshift=0.75cm,xshift=-4.2cm]R1.north)--([yshift=0.75cm,xshift=2cm]R2.north) node[above,near end]{$\textbf{V}_\textbf{DD}$};

        \draw (M5.S) node[ground](GND){};
        \draw (M6.S) node[ground](GND){};

    \end{circuitikz}
    \end{minipage}
    \hspace{4mm}
    \raisebox{1cm}{
    \begin{minipage}[b]{0.35\textwidth}
        \centering
        \begin{tcolorbox}[colframe=black,colback=white,boxrule=0.5pt, width=0.8\textwidth,title=\scriptsize\centering\bfseries Homogeneous,
        fonttitle=\small\bfseries\color{black}, coltitle=black, colbacktitle=white]
            \scriptsize
            \begin{tcolorbox}[colframe=black,colback=cyan!30,boxrule=0pt,left=0pt,right=0pt,top=0pt,bottom=0pt]
            \centering
                Autonomous
            \end{tcolorbox}
            \begin{tcolorbox}[colframe=black,colback=yellow!30,boxrule=0pt,left=0pt,right=0pt,top=0pt,bottom=0pt]
            \centering
                Noise-bandwidth tradeoff 
            \end{tcolorbox}
            \begin{tcolorbox}[colframe=black,colback=brown!30,boxrule=0pt,left=0pt,right=0pt,top=0pt,bottom=0pt]
            \centering
                Noise-power tradeoff 
            \end{tcolorbox}
        \end{tcolorbox}
    \end{minipage}
    }
    \vspace{10pt}
    \caption{Voltage-Controlled Oscillator.}
    \label{fig:circuit:vco}
\end{figure}
\begin{figure}[!htb]
    \centering
    \begin{minipage}[b]{1\textwidth}
     \centering
        \begin{circuitikz}[scale=0.7, transform shape]
        \ctikzset{tripoles/mos style/arrows}
        \ctikzset{resistors/scale=0.5}
        \ctikzset{inductors/scale=0.8, inductors/coils=4}
        \ctikzset{transistors/scale=0.8}
        \ctikzset{quadpoles/transformer/inner=1, quadpoles/transformer/width=1}

        \draw (0,0) node[nmos,anchor=G,rotate=-90](M1){};
        \draw ([yshift=-1.4cm]M1.G) node[anchor=south] {$\text{W}_\text{N1}$};

        \draw (2.2,0) node[nmos,anchor=G,rotate=-90](M2){};
        \draw ([yshift=-1.4cm]M2.G) node[anchor=south] {$\text{W}_\text{N1}$};
        \draw (M2.G) to[short,-*] ($(M2.G)+(0,0.1)$) node[above] {$\text{V}_\text{b1}$};

        \draw (0,-3) node[nmos,anchor=G,rotate=90,yscale=-1](M3){};
        \draw ([yshift=0.8cm]M3.G) node[anchor=south] {$\text{W}_\text{N1}$};

        \draw (2.2,-3) node[nmos,anchor=G,rotate=90,yscale=-1](M4){};
        \draw ([yshift=0.8cm]M4.G) node[anchor=south] {$\text{W}_\text{N1}$};
        \draw (M4.G) to[short,-*] ($(M4.G)+(0,-0.1)$) node[below] {$\text{V}_\text{b1}$};

        \draw (6.12,0) node[nmos,anchor=G,rotate=-90](M5){};
        \draw ([yshift=-1.4cm]M5.G) node[anchor=south] {$\text{W}_\text{N2}$};

        \draw (8.32,0) node[nmos,anchor=G,rotate=-90](M6){};
        \draw ([yshift=-1.4cm]M6.G) node[anchor=south] {$\text{W}_\text{N2}$};
        \draw (M6.G) to[short,-*] ($(M6.G)+(0,0.1)$) node[above] {$\text{V}_\text{b2}$};

        \draw (6.12,-3) node[nmos,anchor=G,rotate=90,yscale=-1](M7){};
        \draw ([yshift=0.8cm]M7.G) node[anchor=south] {$\text{W}_\text{N2}$};

        \draw (8.32,-3) node[nmos,anchor=G,rotate=90,yscale=-1](M8){};
        \draw ([yshift=0.8cm]M8.G) node[anchor=south] {$\text{W}_\text{N2}$};
        \draw (M8.G) to[short,-*] ($(M8.G)+(0,-0.1)$) node[below] {$\text{V}_\text{b2}$};

        \draw (M1.D) to[L,l=$\text{L}_\text{m}$,name=L1] ++(1,0);
        \draw (M3.D) to[L,l_=$\text{L}_\text{m}$,name=L1,mirror] ++(1,0);
        \draw (M5.D) to[L,l=$\text{L}_\text{m}$,name=L1] ++(1,0);
        \draw (M7.D) to[L,l_=$\text{L}_\text{m}$,name=L1,mirror] ++(1,0);
        
        \draw ([xshift=-1.4cm,yshift=0.15cm]M1.G) node[transformer,anchor=B1](T1){};
        \draw (T1.inner dot A1) node[circ]{};
        \draw (T1.inner dot B1) node[circ]{};
        \draw [>=latex, <->] (-2.35,0.1) to[bend left=45] (-1.55,0.1);
        \draw (T1.B1) -- ([xshift=1.4cm]T1.B1);
        \draw ([xshift=1.4cm]T1.B1) -- (M1.G);
        \draw ([xshift=0.05cm]T1-L1.midtap) node[anchor=east] {$\text{L}_\text{ip}$};
        \draw ([xshift=-0.05cm]T1-L2.midtap) node[anchor=west] {$\text{L}_\text{is}$};
        \draw (T1.B2) node[anchor=south,rotate=90] {$\text{V}_\text{b3}$};
        \draw (T1.A1) to[short,-o] ($(T1.A1)+(-1,0)$) node[left] {$\text{V}_\text{in}\text{+}$};

        \draw 
        (T1.B2) node[transformer,anchor=B1](T2){};
        \draw (T2.inner dot A2) node[circ]{};
        \draw (T2.inner dot B2) node[circ]{};
        \draw [>=latex, <->] (-2.35,-3.1) to[bend right=45] (-1.55,-3.1);
        \draw (T2.B2) -- ([xshift=1.4cm]T2.B2);
        \draw ([xshift=1.4cm]T2.B2) -- (M3.G);
        \draw ([xshift=0.05cm]T2-L1.midtap) node[anchor=east] {$\text{L}_\text{ip}$};
        \draw ([xshift=-0.05cm]T2-L2.midtap) node[anchor=west] {$\text{L}_\text{is}$};
        \draw (T2.A2) to[short,-o] ($(T2.A2)+(-1,0)$) node[left] {$\text{V}_\text{in}\text{-}$};

        \draw ([xshift=1.4cm,yshift=0.15cm]M2.G) node[transformer,anchor=A1](T3){};
        \draw (T3.inner dot A1) node[circ]{};
        \draw (T3.inner dot B1) node[circ]{};
        \draw [>=latex, <->] (3.76,0.1) to[bend left=45] (4.56,0.1);
        \draw (T3.A1) -- ([xshift=-0.78cm]T3.A1);
        \draw ([xshift=-0.78cm]T3.A1) -- ([xshift=-0.78cm,yshift=-0.94cm]T3.A1);
        \draw (T3.B1) -- ([xshift=1.4cm]T3.B1);
        \draw ([xshift=1.4cm]T3.B1) -- (M5.G);
        \draw (T3.A2) node[anchor=south,rotate=90] {$\text{V}_\text{DD}$};
        \draw (T3.B2) node[anchor=south,rotate=90] {$\text{V}_\text{b4}$};

        \draw 
        (T3.A2) node[transformer,anchor=A1](T4){};
        \draw (T4.inner dot A2) node[circ]{};
        \draw (T4.inner dot B2) node[circ]{};
        \draw [>=latex, <->] (3.76,-3.1) to[bend right=45] (4.56,-3.1);
        \draw (T4.A2) -- ([xshift=-0.78cm]T4.A2);
        \draw ([xshift=-0.78cm]T4.A2) -- ([xshift=-0.78cm,yshift=1cm]T4.A2);
        \draw (T4.B2) -- ([xshift=1.4cm]T4.B2);
        \draw ([xshift=1.4cm]T4.B2) -- (M7.G);

        \draw ([xshift=1.4cm,yshift=0.15cm]M6.G) node[transformer,anchor=A1](T5){};
        \draw (T5.inner dot A1) node[circ]{};
        \draw (T5.inner dot B1) node[circ]{};
        \draw [>=latex, <->] (9.88,0.1) to[bend left=45] (10.68,0.1);
        \draw (T5.A1) -- ([xshift=-0.78cm]T5.A1);
        \draw ([xshift=-0.78cm]T5.A1) -- ([xshift=-0.78cm,yshift=-0.94cm]T5.A1);
        \draw ([xshift=0.1cm]T5-L1.midtap) node[anchor=east] {$\text{L}_\text{op}$};
        \draw ([xshift=-0.05cm]T5-L2.midtap) node[anchor=west] {$\text{L}_\text{os}$};
        \draw (T5.A2) node[anchor=south,rotate=90] {$\text{V}_\text{DD}$};
        \draw (T5.B1) to[short,-o] ($(T5.B1)+(1,0)$) node[right] {$\text{V}_\text{out}\text{+}$};

        \draw 
        (T5.A2) node[transformer,anchor=A1](T6){};
        \draw (T6.inner dot A2) node[circ]{};
        \draw (T6.inner dot B2) node[circ]{};
        \draw [>=latex, <->] (9.88,-3.1) to[bend right=45] (10.68,-3.1);
        \draw (T6.A2) -- ([xshift=-0.78cm]T6.A2);
        \draw ([xshift=-0.78cm]T6.A2) -- ([xshift=-0.78cm,yshift=1cm]T6.A2);
        \draw ([xshift=0.1cm]T6-L1.midtap) node[anchor=east] {$\text{L}_\text{op}$};
        \draw ([xshift=-0.05cm]T6-L2.midtap) node[anchor=west] {$\text{L}_\text{os}$};
        \draw (T6.B2) to[short,-o] ($(T6.B2)+(1,0)$) node[right] {$\text{V}_\text{out}\text{-}$};

        \draw (M1.S) node[ground](GND){};
        \draw (M3.S) node[ground,yscale=-1](GND){};
        \draw (M5.S) node[ground](GND){};
        \draw (M7.S) node[ground,yscale=-1](GND){};

    \end{circuitikz}
    \end{minipage}\\ \vspace{10pt}
    \begin{minipage}[b]{0.6\textwidth}
        \centering
        \begin{tcolorbox}[colframe=black,colback=white,boxrule=0.5pt, width=0.8\textwidth,title=\scriptsize\centering\bfseries Homogeneous,
        fonttitle=\small\bfseries\color{black}, coltitle=black, colbacktitle=white]
            \scriptsize
            \begin{minipage}{0.48\textwidth}
                \begin{tcolorbox}[colframe=black,colback=red!30,boxrule=0pt,left=0pt,right=0pt,top=0pt,bottom=0pt]
                \centering
                    Gain-bandwidth tradeoff
                \end{tcolorbox}
            \end{minipage}
            \hfill
            \begin{minipage}{0.48\textwidth}
                \begin{tcolorbox}[colframe=black,colback=blue!30,boxrule=0pt,left=0pt,right=0pt,top=0pt,bottom=0pt]
                \centering
                    Gain-power tradeoff
                \end{tcolorbox}
            \end{minipage}
            \begin{tcolorbox}[colframe=black,colback=lime!30,boxrule=0pt,left=0pt,right=0pt,top=0pt,bottom=0pt]
            \centering
                Gain-efficiency tradeoff
            \end{tcolorbox}
        \end{tcolorbox}
    \end{minipage}
    \vspace{10pt}
    \caption{Power Amplifier.}
    \label{fig:circuit:pa}
\end{figure}
\input{./fig/TransmitterSchematic}
\input{./fig/ReceiverSchematic}
\captionsetup[table]{skip=8pt,justification=centering}
\begin{table}[!htb]
    \centering
    \noindent\resizebox{\textwidth}{!}{
       \begin{tabular}{l|l|cc}
        \toprule
         \textbf{Heterogeneous Circuits} & \textbf{Individual Block} & \textbf{Parameter} & \textbf{Sweeping Range} \\ \midrule
         \multirow{12}{*}{\makecell[l]{Transmitter System \\ \textbf{specs}: \\ dc power | bandwidth | output power \\ 
         | voltage swing}} & \multirow{6}{*}{\makecell[l]{Voltage-Controlled Oscillator (VCO) \\ \textbf{specs}: \\ phase noise | tuning range}}& $\text{C}$ &  $\text{50:50:150 (fF)}$ \\
         & & $\text{L}$ & $\text{60:60:180 (pH)}$ \\ 
         & & $\text{R}_{\text{p}}$ & $\text{300:100:500 (\(\Omega\))}$ \\ 
         & & $\text{W}_{\text{N1}}$ & $\text{7.5:2.5:12.5 (\(\upmu\)m)}$ \\ 
         & & $\text{W}_{\text{N2}}$ & $\text{187.5:12.5:212.5 (\(\upmu\)m)}$ \\ 
         & & $\text{W}_{\text{var}}$ & $\text{70:10:90 (\(\upmu\)m)}$ \\ 
         \cmidrule{2-4}
         & \multirow{6}{*}{\makecell[l]{Power Amplifier (PA)  \\ \textbf{specs}: \\ power gain | drain efficiency | PAE}} & $\text{L}_{\text{ip}}$ & $\text{175:175:350 (pH)}$ \\
         & & $\text{L}_{\text{is}}$ & $\text{60:60:120 (pH)}$ \\ 
         & & $\text{L}_{\text{op}}$ & $\text{360:353:713 (pH)}$ \\ 
         & & $\text{L}_{\text{os}}$ & $\text{45:45:90 (pH)}$ \\
         & & $\text{W}_{\text{N3}}$ & $\text{22:5:32 (\(\upmu\)m)}$ \\ 
         & & $\text{W}_{\text{N4}}$ & $\text{16:5:26 (\(\upmu\)m)}$ \\ \midrule

         \multirow{14}{*}{\makecell[l]{Receiver System  \\ \textbf{specs}: \\ dc power | gain | noise figure}}& \multirow{7}{*}{\makecell[l]{Low-Noise Amplifier (LNA)  \\ \textbf{specs}: \\ power gain | $\text{S}_{\text{11}}$ | noise figure}}& $\text{C}_{\text{1}}$ & $\text{130:50:180 (fF)}$ \\
         & & $\text{C}_{\text{2}}$ & $\text{170:50:220 (fF)}$ \\ 
         & & $\text{L}_{\text{d}}$ & $\text{180:50:230 (pH)}$ \\ 
         & & $\text{L}_{\text{g}}$ & $\text{850:100:950 (pH)}$ \\ 
         & & $\text{L}_{\text{s}}$ & $\text{80:10:90 (pH)}$ \\ 
         & & $\text{W}_{\text{N1}}$ & $\text{20:3:26 (\(\upmu\)m)}$ \\ 
         & & $\text{W}_{\text{N2}}$ & $\text{37.5:2.5:42.5 (\(\upmu\)m)}$ \\ 
         \cmidrule{2-4}
         & \multirow{4}{*}{\makecell[l]{Mixer \\ \textbf{specs}: \\ voltage swing | conversion gain}} & $\text{C}_\text{3}$ & $\text{1:0.1:1.1 (pF)}$ \\
         & & $\text{R}_\text{1}$ & $\text{400:100:500 (\(\Omega\))}$ \\ 
         & & $\text{W}_{\text{N3}}$ & $\text{14:2:18 (\(\upmu\)m)}$ \\ 
         & & $\text{W}_{\text{N4}}$ & $\text{6:2:10 (\(\upmu\)m)}$ \\
         \cmidrule{2-4}
         & \multirow{3}{*}{\makecell[l]{Cascode Voltage Amplifier (CVA) \\ \textbf{specs}: \\ gain}} & $\text{R}_{\text{2}}$ & $\text{300:100:400 (\(\Omega\))}$ \\
         & & $\text{W}_{\text{N5}}$ & $\text{26:2:30 (\(\upmu\)m)}$ \\ 
         & & $\text{W}_{\text{N6}}$ & $\text{14:2:18 (\(\upmu\)m)}$ \\ 
         \bottomrule
      \end{tabular}
    }
    \caption{Heterogeneous circuits and the chosen parameters for each block. The sweeping range of selected design parameters is written in the form of $[\texttt{beg}, \texttt{increment}, \texttt{end}]$. Detailed circuit topology is provided in Appendix \ref{appx:circuit}. Specs of each circuit are explained in Appendix \ref{appx:specs}.}
    \label{tab:circuitparam:complex}
\end{table}
\begin{table}[!htb]
    \centering
    \small
    \noindent\resizebox{\textwidth}{!}{
    \begin{tabular}{l|ccc}
        \toprule
        \textbf{Homogeneous Circuit} & \textbf{Parameter} & \textbf{Description} & \textbf{Sweeping Range} \\ \midrule
        \multirow{4}{*}{\makecell[l]{Common-Source Voltage Amplifier (CSVA) \\ \textbf{specs}: \\ dc power | bandwidth | gain}} & $\text{V}_{\text{DD}}$ & $\text{supply voltage}$ & $\text{1.2:0.1:1.8 (V)}$ \\
                             & $\text{V}_{\text{gate}}$ & $\text{gate voltage}$ & $\text{0.6:0.05:0.9 (V)}$\\
                             & $\text{R}_{\text{D}}$& $\text{load resistor}$ & $\text{0.5:0.1:3 (k\(\Omega\))}$\\
                             & $\text{W}_{\text{N}}$& $\text{width of nmos}$ & $\text{3:1:10 (\(\upmu\)m)}$\\ \midrule
        \multirow{4}{*}{\makecell[l]{Cascode Voltage Amplifier (CVA) \\ \textbf{specs}: \\ dc power | bandwidth | gain}} & $\text{R}_{\text{D}}$& $\text{load resistor}$ & $\text{0.5:0.1:2 (k\(\Omega\))}$\\
           & $\text{W}_{\text{N1}}$& $\text{width of nmos}$ & $\text{6:1:17 (\(\upmu\)m)}$\\
           & $\text{W}_{\text{N2}}$& $\text{width of nmos}$ & $\text{5:1:12 (\(\upmu\)m)}$\\
           & $\text{W}_{\text{N3}}$& $\text{width of nmos}$ & $\text{4.5:0.5:9 (\(\upmu\)m)}$\\ \midrule
        \multirow{6}{*}{\makecell[l]{Two-Stage Voltage Amplifier (TSVA) \\ \textbf{specs}: \\ dc power | bandwidth | gain}} & $\text{C}_{\text{1}}$& $\text{miller capacitor}$ & $\text{150:50:250 (fF)}$\\
           & $\text{W}_{\text{P1}}$& $\text{width of pmos}$ & $\text{10:1:18 (\(\upmu\)m)}$\\
           & $\text{W}_{\text{P2}}$& $\text{width of pmos}$ & $\text{7.5:5:22.5 (\(\upmu\)m)}$\\
           & $\text{W}_{\text{N1}}$& $\text{width of nmos}$ & $\text{10:1:18 (\(\upmu\)m)}$\\
           & $\text{W}_{\text{N2}}$& $\text{width of nmos}$ & $\text{7.5:5:22.5 (\(\upmu\)m)}$\\
           & $\text{W}_{\text{N3}}$& $\text{width of nmos}$ & $\text{16:2:24 (\(\upmu\)m)}$\\ \midrule
        \multirow{7}{*}{\makecell[l]{Low-Noise Amplifier (LNA) \\ \textbf{specs}: \\ dc power | bandwidth | power gain \\ | $\text{S}_{\text{11}}$
        | noise figure}} & $\text{C}_{\text{1}}$& $\text{output capacitor}$ & $\text{300:100:600 (fF)}$\\
           & $\text{C}_{\text{2}}$& $\text{input capacitor}$ & $\text{200:100:500 (fF)}$\\
           & $\text{L}_{\text{d}}$& $\text{drain inductor}$ & $\text{3:1:5 (nH)}$\\
           & $\text{L}_{\text{g}}$& $\text{gate inductor}$ & $\text{8.4:1:11.4 (nH)}$\\
           & $\text{L}_{\text{s}}$& $\text{source inductor}$ & $\text{0.6:0.1:0.8 (nH)}$\\
           & $\text{W}_{\text{N1}}$& $\text{width of nmos}$ & $\text{25:1.25:30 (\(\upmu\)m)}$\\ 
           & $\text{W}_{\text{N2}}$& $\text{width of nmos}$ & $\text{25:1.25:30 (\(\upmu\)m)}$\\ \midrule
        \multirow{4}{*}{\makecell[l]{Mixer \\ \textbf{specs}: \\ dc power | voltage swing \\ | conversion gain | noise figure}} & $\text{C}$& $\text{coupling capacitor}$ & $\text{0.5:0.1:1.5 (pF)}$\\
           & $\text{R}$& $\text{load resistor}$ & $\text{200:25:500 (\(\Omega\))}$\\
           & $\text{W}_{\text{N1}}$& $\text{width of nmos}$ & $\text{15:1:25 (\(\upmu\)m)}$\\ 
           & $\text{W}_{\text{N2}}$& $\text{width of nmos}$ & $\text{5:1:15 (\(\upmu\)m)}$\\ \midrule
       \multirow{7}{*}{\makecell[l]{Voltage-Controlled Oscillator (VCO) \\ \textbf{specs}: \\ dc power | frequency | phase noise \\ | tuning range}} & $\text{C}$& $\text{capacitor in resonant tank}$& $\text{100:25:200 (fF)}$\\
           & $\text{L}$& $\text{inductor in resonant tank}$& $\text{2:1:6 (nH)}$\\
           & $\text{R}_{\text{p}}$& $\text{parallel resistor}$& $\text{1:1:4 (k\(\Omega\))}$\\
           & $\text{W}_{\text{N1}}$& $\text{width of nmos}$& $\text{24:8:56 (\(\upmu\)m)}$\\ 
           & $\text{W}_{\text{N2}}$& $\text{width of nmos}$& $\text{11:1:12 (\(\upmu\)m)}$\\ 
           & $\text{W}_{\text{N3}}$& $\text{width of nmos}$& $\text{96:32:160 (\(\upmu\)m)}$\\ 
           & $\text{W}_{\text{var}}$& $\text{width of nmos capacitor}$& $\text{75:12.5:125 (\(\upmu\)m)}$\\ \midrule
        \multirow{7}{*}{\makecell[l]{Power Amplifier (PA) \\ \textbf{specs}: \\ dc power | $\text{S}_{\text{11}}$ | $\text{S}_{\text{22}}$ | power gain \\ | PAE | drain efficiency | $\text{P}_{\text{sat}}$}} & $\text{L}_{\text{ip}}$& $\text{input primary inductor}$& $\text{175:175:525 (pH)}$\\
           & $\text{L}_{\text{is}}$& $\text{input secondary inductor}$& $\text{40:40:120 (pH)}$\\
           & $\text{L}_{\text{m}}$& $\text{inter-stage matching inductor}$& $\text{87.5:87.5:263 (pH)}$\\
           & $\text{L}_{\text{op}}$& $\text{output primary inductor}$& $\text{238:238:714 (pH)}$\\ 
           & $\text{L}_{\text{os}}$& $\text{output secondary inductor}$& $\text{30:30:90 (pH)}$\\
           & $\text{W}_{\text{N1}}$& $\text{width of nmos}$& $\text{16:3:28 (\(\upmu\)m)}$\\ 
           & $\text{W}_{\text{N2}}$& $\text{width of nmos}$& $\text{24:4:40 (\(\upmu\)m)}$\\ 
         \bottomrule
    \end{tabular}
    }
    \caption{Homogeneous circuits and the chosen parameters for each circuit. The sweeping range of selected design parameters is written in the form of $[\texttt{beg}, \texttt{increment}, \texttt{end}]$. Detailed circuit topology is provided in Appendix \ref{appx:circuit}. Specifications of each circuit are explained in Appendix \ref{appx:specs}.}
    \label{tab:circuitparam:basic}
\end{table} 
\clearpage
\section{Model Details}\label{appx:model}
\captionsetup[table]{skip=8pt,justification=centering}
\begin{table}[!htb]
    \centering
    \noindent\resizebox{\textwidth}{!}{
    \begin{tabular}{l|ccc}
        \toprule
        \textbf{Model} & \textbf{Parameter} & \textbf{Description} & \textbf{Value} \\ \midrule
        \multirow{6}{*}{\makecell[l]{Transformer}} & $\texttt{dim\_model}$ & $\text{first fully connected layer dim}$ & $\text{200}$ \\
                             & $\texttt{num\_heads}$ & $\text{heads in the multi-head attention models}$ & $\text{2}$\\
                             & $\texttt{dim\_hidden}$& $\text{load resistor}$ & $\text{200}$\\
                             & $\texttt{dropout\_p}$& $\text{the dropout probability}$ & $\text{0.1}$\\
                             & $\texttt{num\_encoder\_layers}$& $\text{number of layers in transformer encoder}$ & $\text{6}$\\
                             & $\texttt{activation}$& $\text{activation function of transformer encoder}$ & $\text{relu}$\\ \midrule
        \multirow{2}{*}{\makecell[l]{Multi Layer Perception (MLP)}} & $\texttt{num\_layers}$& $\text{number of fully connected layers}$ & $\text{7}$\\
           & $\texttt{dim\_layers}$& $\text{dimension of layers}$ & $\text{[200, 300, 500, 500, 300, 200]}$\\ \midrule
        \multirow{2}{*}{\makecell[l]{Support Vector Regressor (SVR)}} & $\texttt{kernel}$& $\text{kernel type of the algorithm}$ & $\text{rbf}$\\
           & $\texttt{multi\_target\_regression\_type}$& $\text{type of combining multiple SVRs}$ & $\text{MultiOutputRegression}$\\
          \midrule
        \multirow{2}{*}{\makecell[l]{K Nearest Neighbors Regressor (KNN)}} & $\texttt{n\_neighbors}$& $\text{number of neighbors}$ & $\text{5}$\\
           & $\texttt{weights}$& $\text{weight function used in prediction}$ & $\text{uniform}$\\
            \midrule
        \multirow{2}{*}{\makecell[l]{Random Forest Regressor (RF)}} & $\texttt{n\_estimators}$& $\text{number of trees in the forest}$ & $\text{100}$\\
           & $\texttt{criterion}$& $\text{function to measure the quality of a split}$ & $\text{squared\_error ($l_2$ Loss)}$\\
         \bottomrule
    \end{tabular}

    }
    \caption{Models and the chosen parameters for each model.}
    \label{tab:modelparam}
\end{table}

\captionsetup[table]{skip=8pt,justification=centering}
\begin{table}[!htb]
    \centering
    \noindent\resizebox{\textwidth}{!}{
    \begin{tabular}{l|ccccccccc}
        \toprule
        \textbf{Circuit} & \text{CSVA} & \text{CVA} & \text{TSVA} & \text{LNA} & \text{Mixer} & \text{VCO} & \text{PA} & \text{Transmitter} & \text{Receiver} \\ \midrule
        \multirow{1}{*}{\makecell[c]{\textbf{Dataset Size}}} & $\text{7.8k}$ & $\text{15.1k}$ & $\text{19.3k}$ & $\text{32k}$ & $\text{17.1k}$ & $\text{13.5k}$ & $\text{5.6k}$ & $\text{95.3k}$ & $\text{155.4k}$\\
         \bottomrule
    \end{tabular}

    }
    \caption{Number of data points for each circuit.}
    \label{tab:modelparam2}
\end{table}

\section{Specifications of Each Circuit}\label{appx:specs}
In this section, we review the definitions of various performance metrics that are simulated for the homogeneous and heterogeneous circuits in this work.\\
\textbf{Voltage gain} of an amplifier, denoted commonly as $A_v$, is the ratio of the output amplified voltage $V_{out}$ to the input voltage $V_{in}$ in an amplifier circuit. \begin{equation}
   A_v = \frac{V_{out}}{V_{in}}
   \end{equation}
   \textbf{Bandwidth} is the range of frequencies over which an amplifier can operate effectively, defined by the difference between the upper and lower cutoff frequencies.
   \begin{equation}
   BW = f_{high} - f_{low}
   \end{equation}
   For amplifiers with low-pass profile, the bandwidth is defined as the frequency at which the dB amount of voltage gain drops from the low-frequency gain by 3 dB.\\
   The total power that the circuit draws from the power supply is known as \textbf{power consumption}, and it is calculated as the product of supply voltage and supply current.
     \begin{equation}
   P = V_{supply} \times I_{supply}
   \end{equation}
   \textbf{Conversion gain} is the measure of the signal amplification in a mixer, expressed in decibels, comparing the output signal to the input signal.
     \begin{equation}
   CG = 20 \log \left( \frac{V_{out}}{V_{in}} \right) \, \text{dB}
   \end{equation}
   \textbf{Noise figure} is the ratio of the input signal-to-noise ratio to the output signal-to-noise ratio, expressed in decibels, indicating the noise performance of a low-noise amplifier, mixer, or receiver chain.
    \begin{equation}
   NF = 10 \log \left( \frac{SNR_{in}}{SNR_{out}} \right) \, \text{dB}
   \end{equation}
   Intermediate frequency (IF) \textbf{voltage swing} refers to the peak-to-peak voltage of the mixer's intermediate frequency signal.
 
   The \textbf{oscillation frequency} is the frequency at which the oscillator produces its periodic signal, which is normally controlled by the circuit's inductance and capacitance in an LC-based oscillator.
\begin{equation}
   f_{\text{osc}} = \frac{1}{2 \pi \sqrt{LC}}
   \end{equation}
   The power that the oscillator provides to a given load is known as \textbf{output power}, and it is computed by dividing the root-mean-square (RMS) output voltage squared by the load resistance.
\begin{equation}
   P_{\text{out}} = \frac{V_{\text{out,RMS}}^2}{R_{\text{load}}}
   \end{equation}
   \textbf{Phase noise} is a measure of the oscillator's frequency stability, representing the noise power in a 1 Hz bandwidth at a specific offset frequency from the carrier, relative to the carrier power, where $S_{\phi}(f)$ is the phase noise power spectral density.
    \begin{equation}
   L(f) = 10 \log \left( \frac{S_{\phi}(f)}{2 P_{\text{carrier}}} \right) \, \text{dBc/Hz}
   \end{equation}
   The oscillator's \textbf{tuning range} is the range of frequencies that it can be adjusted over, measured from maximum to minimum.
\begin{equation}
   TR = f_{\text{max}} - f_{\text{min}}
   \end{equation}
   $\textbf{S}_\textbf{11}$ represents the ratio of the reflected voltage wave to the incident voltage wave at the power amplifier's input, which shows how much of the input signal is reflected.
\begin{equation}
   S_{11} = \frac{V_{reflected,1}}{V_{incident,1}}
   \end{equation}
   $\textbf{S}_\textbf{22}$ represents the ratio of the reflected voltage wave to the incident voltage wave at the power amplifier's output, which shows how much of the output signal is reflected.
     \begin{equation}
   S_{22} = \frac{V_{reflected,2}}{V_{incident,2}}
   \end{equation}
   \textbf{Large signal gain} is the ratio of the output power to the input power of the amplifier under large signal conditions.
   \begin{equation}
   G_{\text{LS}} = 20 \log \left( \frac{P_{out}}{P_{in}} \right) \, \text{dB}
   \end{equation}
   \textbf{Power added efficiency} is the amplifier's efficiency in converting DC power into radio frequency output power while taking input power into account.
\begin{equation}
   PAE = \frac{P_{out} - P_{in}}{P_{DC}} \times 100\%
   \end{equation}
   \textbf{Drain efficiency} is defined as the ratio of radio frequency output power to total DC power consumed by the amplifier.
\begin{equation}
   DE = \frac{P_{out}}{P_{DC}} \times 100\%
   \end{equation}
  \textbf{Saturated power} 
or Psat denotes the highest possible output power level that a power amplifier can produce upon reaching saturation. Up until this point, increasing the input power further does not significantly increase the output power.
\begin{equation}
P_{\text{sat}} (\text{dBm}) = 10 \log_{10} \left( \frac{P_{\text{sat}} (\text{mW})}{1 \text{mW}} \right)
\end{equation}
  \textbf{Power gain} 
 is a measure of how much a circuit increases the power of a signal from its input to its output. 
\begin{equation}
G_P = \frac{P_{\text{out}}}{P_{\text{in}}}
\end{equation}
    \\ \textbf{Voltage-controlled oscillator's (VCO) output power} is the amount of electrical power delivered at its output terminal. The output power \(P_{\text{out}}\) is proportional to the square of the RMS voltage \( V_{\text{rms}} \) and the load resistance \( R_{\text{load}} \).
\begin{equation}
P_{\text{out}} = \frac{1}{2} \cdot V_{\text{rms}}^2 \cdot R_{\text{load}}
\end{equation}

\textbf{Transmitter output power} is the amount of electrical power delivered by the transmitter to the antenna for transmission as electromagnetic waves.
\begin{equation}
P_{\text{out}} = P_{\text{in}} \cdot G_t
\end{equation}
where, $P_{\text{out}}$  is the transmitter output power.
     $P_{\text{in}}$  is the input power to the transmitter, and
    $G_t$  is the gain of the transmitter.
     \\ \textbf{Transducer Gain ( $G_t$ )} of Low Noise Amplifier (LNA) refers to the ratio of the output signal power to the available input signal power which includes the matching effect as well.
     \begin{equation}
G_T = \frac{P_{\text{out}}}{P_{\text{in}}}
\end{equation}

\section{Results on More Circuits}\label{appx:additional}

\textbf{Results on Common-Source Voltage Amplifier (CSVA)}. 

\pgfplotstableread[col sep=comma]{VoltageGainTransformer.csv}{\loadedtableA}
\pgfplotstableread[col sep=comma]{BandwidthTransformer.csv}{\loadedtableB}
\pgfplotstableread[col sep=comma]{PowerConsumptionTransformer.csv}{\loadedtableC}

\pgfplotstableread[col sep=comma]{VoltageGainMLP.csv}{\loadedtableAA}
\pgfplotstableread[col sep=comma]{BandwidthMLP.csv}{\loadedtableBB}
\pgfplotstableread[col sep=comma]{PowerConsumptionMLP.csv}{\loadedtableCC}

\pgfplotstableread[col sep=comma]{VoltageGainSVR.csv}{\loadedtableAAA}
\pgfplotstableread[col sep=comma]{BandwidthSVR.csv}{\loadedtableBBB}
\pgfplotstableread[col sep=comma]{PowerConsumptionSVR.csv}{\loadedtableCCC}

\pgfplotstableread[col sep=comma]{VoltageGainRF.csv}{\loadedtableAAAA}
\pgfplotstableread[col sep=comma]{BandwidthRF.csv}{\loadedtableBBBB}
\pgfplotstableread[col sep=comma]{PowerConsumptionRF.csv}{\loadedtableCCCC}

\begin{figure}[!htb]
    \centering
    \begin{tikzpicture}
        \begin{axis} [
            log origin=infty,
            title={\tiny Voltage Gain}, title style={yshift=-1.5ex},
            ticklabel style={font=\fontsize{4}{5}\selectfont},
            width=0.3\linewidth, height=.25\linewidth,
            xlabel={\tiny relative error (\%)}, x label style={at={(axis description cs:0.5,-0.1)},anchor=north},
            xmax=0.6, xtick={ 0, 0.1, 0.2, 0.3, 0.4, 0.5 }, xticklabels = {0, 10, 20, 30, 40, 50},
            ylabel={\tiny prob density}, ymin=0, y label style={at={(axis description cs:-0.15,.5)},anchor=south},
            ytick={ 1, 5 }, yticklabels={ 1, 5 },
            legend style={at={ (1.0,.6)}, anchor=east, draw=none, fill=none, font=\fontsize{4}{5}\selectfont },
            legend cell align={left}
        ]

            \addplot[thin, fill=blue!20, draw=blue!70, smooth, opacity=0.6] table [x=bin, y=cnt, col sep=comma] {\loadedtableAA} \closedcycle;
            \addlegendentry{MLP}
            
            \addplot[thin, fill=red!20, draw=red!70, smooth, opacity=0.6] table [x=bin, y=cnt, col sep=comma] {\loadedtableA} \closedcycle;
            \addlegendentry{Trans}

            \addplot[thin, fill=brown!20, draw=brown!70, smooth, opacity=0.6] table [x=bin, y=cnt, col sep=comma] {\loadedtableAAAA} \closedcycle;
            \addlegendentry{RF}

            \addplot[thin, fill=green!20, draw=green!70, smooth, opacity=0.4] table [x=bin, y=cnt, col sep=comma] {\loadedtableAAA} \closedcycle;
            \addlegendentry{SVR}

        \end{axis}
    \end{tikzpicture}   
    \begin{tikzpicture}
        \begin{axis} [
            log origin=infty,
            title={\tiny Bandwidth}, title style={yshift=-1.5ex},
            ticklabel style={font=\fontsize{4}{5}\selectfont},
            width=0.3\linewidth, height=.25\linewidth,
            xlabel={\tiny relative error (\%)}, x label style={at={(axis description cs:0.5,-0.1)},anchor=north},
            xtick={ 0, 0.1, 0.2, 0.3, 0.4 }, xticklabels = {0, 10, 20, 30, 40}, scaled x ticks=false,
            xmax=0.5, ymin=0,
            ytick={ 1, 5, 10, 30 }, yticklabels={ 1, 5, 10, 30 },
            legend style={at={ (1.0,.6)}, anchor=east, draw=none, fill=none, font=\fontsize{4}{5}\selectfont },
            legend cell align={left}
        ]    

            \addplot[thin, fill=red!20, draw=red!70, smooth, opacity=0.7] table [x=bin, y=cnt, col sep=comma] {\loadedtableB} \closedcycle;
            \addlegendentry{Trans}
            
            \addplot[thin, fill=blue!20, draw=blue!70, smooth, opacity=0.6] table [x=bin, y=cnt, col sep=comma] {\loadedtableBB} \closedcycle;
            \addlegendentry{MLP}

            \addplot[thin, fill=green!20, draw=green!70, smooth, opacity=0.6] table [x=bin, y=cnt, col sep=comma] {\loadedtableBBB} \closedcycle;
            \addlegendentry{SVR}

            \addplot[thin, fill=brown!20, brown=blue!70, smooth, opacity=0.4] table [x=bin, y=cnt, col sep=comma] {\loadedtableBBBB} \closedcycle;
            \addlegendentry{RF}

        \end{axis}
    \end{tikzpicture} 
    \begin{tikzpicture}
        \begin{axis} [
            log origin=infty,
            title={\tiny Power Consumption}, title style={yshift=-1.5ex},
            ticklabel style={font=\fontsize{4}{5}\selectfont},
            width=0.3\linewidth, height=.25\linewidth,
            xlabel={\tiny relative error (\%)},  x label style={at={(axis description cs:0.5,-0.1)},anchor=north},
            xtick={ 0, 0.1, 0.2, 0.3}, xticklabels = {0, 10, 20, 30}, scaled x ticks=false,
            xmax=0.3, ymin=0, ymax=30,
            ytick={ 10, 20, 30 }, yticklabels={ 10, 20, 30 },
            legend style={at={ (1.0,.6)}, anchor=east, draw=none, fill=none, font=\fontsize{4}{5}\selectfont },
            legend cell align={left}
        ]     

            \addplot[thin, fill=brown!20, brown=blue!70, smooth, opacity=0.6] table [x=bin, y=cnt, col sep=comma] {\loadedtableCCCC} \closedcycle;
            \addlegendentry{RF}
            
            \addplot[thin, fill=blue!20, draw=blue!70, smooth, opacity=0.7] table [x=bin, y=cnt, col sep=comma] {\loadedtableCC} \closedcycle;
            \addlegendentry{MLP}

            \addplot[thin, fill=red!20, draw=red!50, smooth, opacity=0.6] table [x=bin, y=cnt, col sep=comma] {\loadedtableC} \closedcycle;
            \addlegendentry{Trans}

            \addplot[thin, fill=green!20, draw=green!70, smooth, opacity=0.6] table [x=bin, y=cnt, col sep=comma] {\loadedtableCCC} \closedcycle;
            \addlegendentry{SVR}

        \end{axis}
    \end{tikzpicture} 
    \caption{Common-Source Voltage Amplifier}
    \label{fig:error:compare:commonsource}
    \vspace{-3mm}
\end{figure}

\textbf{Results on Cascode Voltage Amplifier (Cascode)}.

\pgfplotstableread[col sep=comma]{VoltageGainTransformer.csv}{\loadedtableA}
\pgfplotstableread[col sep=comma]{BandwidthTransformer.csv}{\loadedtableB}
\pgfplotstableread[col sep=comma]{PowerConsumptionTransformer.csv}{\loadedtableC}

\pgfplotstableread[col sep=comma]{VoltageGainMLP.csv}{\loadedtableAA}
\pgfplotstableread[col sep=comma]{BandwidthMLP.csv}{\loadedtableBB}
\pgfplotstableread[col sep=comma]{PowerConsumptionMLP.csv}{\loadedtableCC}

\pgfplotstableread[col sep=comma]{VoltageGainSVR.csv}{\loadedtableAAA}
\pgfplotstableread[col sep=comma]{BandwidthSVR.csv}{\loadedtableBBB}
\pgfplotstableread[col sep=comma]{PowerConsumptionSVR.csv}{\loadedtableCCC}

\pgfplotstableread[col sep=comma]{VoltageGainRF.csv}{\loadedtableAAAA}
\pgfplotstableread[col sep=comma]{BandwidthRF.csv}{\loadedtableBBBB}
\pgfplotstableread[col sep=comma]{PowerConsumptionRF.csv}{\loadedtableCCCC}

\begin{figure}[!htb]
    \centering
    \begin{tikzpicture}
        \begin{axis} [
            log origin=infty,
            title={\tiny Voltage Gain}, title style={yshift=-1.5ex},
            ticklabel style={font=\fontsize{4}{5}\selectfont},
            width=0.3\linewidth, height=.25\linewidth,
            xlabel={\tiny relative error (\%)}, x label style={at={(axis description cs:0.5,-0.1)},anchor=north},
            xmax=0.6, xtick={ 0, 0.1, 0.2, 0.3, 0.4, 0.5 }, xticklabels = {0, 10, 20, 30, 40, 50},
            ylabel={\tiny prob density}, ymin=0, y label style={at={(axis description cs:-0.15,.5)},anchor=south},
            ytick={ 1, 5 }, yticklabels={ 1, 5 },
            legend style={at={ (1.0,.6)}, anchor=east, draw=none, fill=none, font=\fontsize{4}{5}\selectfont },
            legend cell align={left}
        ]
        
            \addplot[thin, fill=blue!20, draw=blue!70, smooth, opacity=0.6] table [x=bin, y=cnt, col sep=comma] {\loadedtableAA} \closedcycle;
            \addlegendentry{MLP}
            
            \addplot[thin, fill=red!20, draw=red!70, smooth, opacity=0.6] table [x=bin, y=cnt, col sep=comma] {\loadedtableA} \closedcycle;
            \addlegendentry{Trans}

            \addplot[thin, fill=brown!20, draw=brown!70, smooth, opacity=0.6] table [x=bin, y=cnt, col sep=comma] {\loadedtableAAAA} \closedcycle;
            \addlegendentry{RF}

            \addplot[thin, fill=green!20, draw=green!70, smooth, opacity=0.4] table [x=bin, y=cnt, col sep=comma] {\loadedtableAAA} \closedcycle;
            \addlegendentry{SVR}

        \end{axis}
    \end{tikzpicture}   
    \begin{tikzpicture}
        \begin{axis} [
            log origin=infty,
            title={\tiny Bandwidth}, title style={yshift=-1.5ex},
            ticklabel style={font=\fontsize{4}{5}\selectfont},
            width=0.3\linewidth, height=.25\linewidth,
            xlabel={\tiny relative error (\%)}, x label style={at={(axis description cs:0.5,-0.1)},anchor=north},
            xtick={ 0, 0.1, 0.2, 0.3}, xticklabels = {0, 10, 20, 30}, scaled x ticks=false,
            xmax=0.35, ymin=0,
            ytick={ 1, 5, 10, 30 }, yticklabels={ 1, 5, 10, 30 },
            legend style={at={ (1.0,.6)}, anchor=east, draw=none, fill=none, font=\fontsize{4}{5}\selectfont },
            legend cell align={left}
        ]    

            \addplot[thin, fill=green!20, draw=green!70, smooth, opacity=0.6] table [x=bin, y=cnt, col sep=comma] {\loadedtableBBB} \closedcycle;
            \addlegendentry{SVR}
            
            \addplot[thin, fill=blue!20, draw=blue!70, smooth, opacity=0.6] table [x=bin, y=cnt, col sep=comma] {\loadedtableBB} \closedcycle;
            \addlegendentry{MLP}

            \addplot[thin, fill=brown!20, brown=blue!70, smooth, opacity=0.5] table [x=bin, y=cnt, col sep=comma] {\loadedtableBBBB} \closedcycle;
            \addlegendentry{RF}

            \addplot[thin, fill=red!20, draw=red!70, smooth, opacity=0.5] table [x=bin, y=cnt, col sep=comma] {\loadedtableB} \closedcycle;
            \addlegendentry{Trans}

        \end{axis}
    \end{tikzpicture} 
    \begin{tikzpicture}
        \begin{axis} [
            log origin=infty,
            title={\tiny Power Consumption}, title style={yshift=-1.5ex},
            ticklabel style={font=\fontsize{4}{5}\selectfont},
            width=0.3\linewidth, height=.25\linewidth,
            xlabel={\tiny relative error (\%)},  x label style={at={(axis description cs:0.5,-0.1)},anchor=north},
            xtick={ 0, .05, .1, .15}, xticklabels = {0, 5, 10, 15}, scaled x ticks=false,
            xmax=0.2, ymin=0, ymax=30,
            ytick={ 10, 20, 30 }, yticklabels={ 10, 20, 30 },
            legend style={at={ (1.0,.6)}, anchor=east, draw=none, fill=none, font=\fontsize{4}{5}\selectfont },
            legend cell align={left}
        ]      

            \addplot[thin, fill=green!20, draw=green!70, smooth, opacity=0.6] table [x=bin, y=cnt, col sep=comma] {\loadedtableCCC} \closedcycle;
            \addlegendentry{SVR}

            \addplot[thin, fill=brown!20, brown=blue!70, smooth, opacity=0.6] table [x=bin, y=cnt, col sep=comma] {\loadedtableCCCC} \closedcycle;
            \addlegendentry{RF}

            \addplot[thin, fill=red!20, draw=red!50, smooth, opacity=0.6] table [x=bin, y=cnt, col sep=comma] {\loadedtableC} \closedcycle;
            \addlegendentry{Trans}

            \addplot[thin, fill=blue!20, draw=blue!70, smooth, opacity=0.7] table [x=bin, y=cnt, col sep=comma] {\loadedtableCC} \closedcycle;
            \addlegendentry{MLP}

        \end{axis}
    \end{tikzpicture} 
    \caption{Cascode Voltage Amplifier}
    \label{fig:error:compare:cascode}
    \vspace{-3mm}
\end{figure}

\textbf{Results on Low-Noise Amplifier (LNA)}.

\pgfplotstableread[col sep=comma]{SOneTransformer.csv}{\loadedtableA}
\pgfplotstableread[col sep=comma]{GTMaxTransformer.csv}{\loadedtableB}
\pgfplotstableread[col sep=comma]{NFMinTransformer.csv}{\loadedtableC}
\pgfplotstableread[col sep=comma]{BandwidthTransformer.csv}{\loadedtableD}
\pgfplotstableread[col sep=comma]{PowerConsumptionTransformer.csv}{\loadedtableE}

\pgfplotstableread[col sep=comma]{SOneMLP.csv}{\loadedtableAA}
\pgfplotstableread[col sep=comma]{GTMaxMLP.csv}{\loadedtableBB}
\pgfplotstableread[col sep=comma]{NFMinMLP.csv}{\loadedtableCC}
\pgfplotstableread[col sep=comma]{BandwidthMLP.csv}{\loadedtableDD}
\pgfplotstableread[col sep=comma]{PowerConsumptionMLP.csv}{\loadedtableEE}

\pgfplotstableread[col sep=comma]{VoltageGainSVR.csv}{\loadedtableAAA}
\pgfplotstableread[col sep=comma]{GTMaxSVR.csv}{\loadedtableBBB}
\pgfplotstableread[col sep=comma]{NFMinSVR.csv}{\loadedtableCCC}
\pgfplotstableread[col sep=comma]{BandwidthSVR.csv}{\loadedtableDDD}
\pgfplotstableread[col sep=comma]{PowerConsumptionSVR.csv}{\loadedtableEEE}

\pgfplotstableread[col sep=comma]{VoltageGainRF.csv}{\loadedtableAAAA}
\pgfplotstableread[col sep=comma]{GTMaxRF.csv}{\loadedtableBBBB}
\pgfplotstableread[col sep=comma]{NFMinRF.csv}{\loadedtableCCCC}
\pgfplotstableread[col sep=comma]{BandwidthRF.csv}{\loadedtableDDDD}
\pgfplotstableread[col sep=comma]{PowerConsumptionRF.csv}{\loadedtableEEEE}

\begin{figure}[!htb]
    \centering
    \begin{tikzpicture}
        \begin{axis} [
            log origin=infty,
            title={\tiny S11}, title style={yshift=-1.5ex},
            ticklabel style={font=\fontsize{4}{5}\selectfont},
            width=0.24\linewidth, height=.25\linewidth,
            xlabel={\tiny relative error (\%)}, x label style={at={(axis description cs:0.5,-0.1)},anchor=north},
            xmax=0.6, xtick={ 0, 0.1, 0.2, 0.3, 0.4, 0.5 }, xticklabels = {0, 10, 20, 30, 40, 50},
            ylabel={\tiny prob density}, ymin=0, y label style={at={(axis description cs:-0.15,.5)},anchor=south},
            ytick={ 1, 5 }, yticklabels={ 1, 5 },
            legend style={at={ (1.0,.6)}, anchor=east, draw=none, fill=none, font=\fontsize{4}{5}\selectfont },
            legend cell align={left}
        ]

            \addplot[thin, fill=green!20, draw=green!70, smooth, opacity=0.4] table [x=bin, y=cnt, col sep=comma] {\loadedtableAAA} \closedcycle;
            \addlegendentry{SVR}
            
            \addplot[thin, fill=brown!20, draw=brown!70, smooth, opacity=0.6] table [x=bin, y=cnt, col sep=comma] {\loadedtableAAAA} \closedcycle;
            \addlegendentry{RF}

            \addplot[thin, fill=blue!20, draw=blue!70, smooth, opacity=0.6] table [x=bin, y=cnt, col sep=comma] {\loadedtableAA} \closedcycle;
            \addlegendentry{MLP}
            
            \addplot[thin, fill=red!20, draw=red!70, smooth, opacity=0.6] table [x=bin, y=cnt, col sep=comma] {\loadedtableA} \closedcycle;
            \addlegendentry{Trans}

        \end{axis}
    \end{tikzpicture} \hspace{-2mm}
    \begin{tikzpicture}
        \begin{semilogyaxis} [
            log origin=infty,
            title={\tiny GTMax}, title style={yshift=-1.5ex},
            ticklabel style={font=\fontsize{4}{5}\selectfont},
            width=0.24\linewidth, height=.25\linewidth,
            xlabel={\tiny relative error (\%)}, x label style={at={(axis description cs:0.5,-0.1)},anchor=north},
            xtick={ 0, .02, .04 }, xticklabels = {0, 2, 4}, scaled x ticks=false,
            xmax=0.10, ymin=0,
            ytick={ 1, 5, 10, 30 }, yticklabels={ 1, 5, 10, 30 },
            legend style={at={ (1.0,.6)}, anchor=east, draw=none, fill=none, font=\fontsize{4}{5}\selectfont },
            legend cell align={left}
        ]    

            \addplot[thin, fill=brown!20, brown=blue!70, smooth, opacity=0.6] table [x=bin, y=cnt, col sep=comma] {\loadedtableBBBB} \closedcycle;
            \addlegendentry{RF}

            \addplot[thin, fill=green!20, draw=green!70, smooth, opacity=0.6] table [x=bin, y=cnt, col sep=comma] {\loadedtableBBB} \closedcycle;
            \addlegendentry{SVR}
            
            \addplot[thin, fill=blue!20, draw=blue!70, smooth, opacity=0.6] table [x=bin, y=cnt, col sep=comma] {\loadedtableBB} \closedcycle;
            \addlegendentry{MLP}

            \addplot[thin, fill=red!20, draw=red!70, smooth, opacity=0.7] table [x=bin, y=cnt, col sep=comma] {\loadedtableB} \closedcycle;
            \addlegendentry{Trans}

        \end{semilogyaxis}
    \end{tikzpicture} \hspace{-2mm}
    \begin{tikzpicture}
        \begin{semilogyaxis} [
            log origin=infty,
            title={\tiny NFMin}, title style={yshift=-1.5ex},
            ticklabel style={font=\fontsize{4}{5}\selectfont},
            width=0.24\linewidth, height=.25\linewidth,
            xlabel={\tiny relative error (\%)},  x label style={at={(axis description cs:0.5,-0.1)},anchor=north},
            xtick={ 0, .02, .04}, xticklabels = {0, 2, 4}, scaled x ticks=false,
            xmax=0.06, ymin=0, ymax=100,
            ytick={ 10, 50, 100 }, yticklabels={ 10, 50, 100 },
            legend style={at={ (1.0,.6)}, anchor=east, draw=none, fill=none, font=\fontsize{4}{5}\selectfont },
            legend cell align={left}
        ]      

            \addplot[thin, fill=green!20, draw=green!70, smooth, opacity=0.6] table [x=bin, y=cnt, col sep=comma] {\loadedtableCCC} \closedcycle;
            \addlegendentry{SVR}

            \addplot[thin, fill=brown!20, brown=blue!70, smooth, opacity=0.6] table [x=bin, y=cnt, col sep=comma] {\loadedtableCCCC} \closedcycle;
            \addlegendentry{RF}

            \addplot[thin, fill=blue!20, draw=blue!70, smooth, opacity=0.7] table [x=bin, y=cnt, col sep=comma] {\loadedtableCC} \closedcycle;
            \addlegendentry{MLP}

            \addplot[thin, fill=red!20, draw=red!50, smooth, opacity=0.6] table [x=bin, y=cnt, col sep=comma] {\loadedtableC} \closedcycle;
            \addlegendentry{Trans}

        \end{semilogyaxis}
    \end{tikzpicture} \hspace{-2mm}
    \begin{tikzpicture}
        \begin{semilogyaxis} [
            log origin=infty,
            title={\tiny Bandwidth}, title style={yshift=-1.5ex},
            ticklabel style={font=\fontsize{4}{5}\selectfont},
            width=0.24\linewidth, height=.25\linewidth,
            xlabel={\tiny relative error (\%)},  x label style={at={(axis description cs:0.5,-0.1)},anchor=north},
            xtick={ 0, .02, .04}, xticklabels = {0, 2, 4}, scaled x ticks=false,
            xmax=0.06, ymin=0, ymax=100,
            ytick={ 10, 50, 100 }, yticklabels={ 10, 50, 100 },
            legend style={at={ (1.0,.6)}, anchor=east, draw=none, fill=none, font=\fontsize{4}{5}\selectfont },
            legend cell align={left}
        ]      

            \addplot[thin, fill=green!20, draw=green!70, smooth, opacity=0.6] table [x=bin, y=cnt, col sep=comma] {\loadedtableDDD} \closedcycle;
            \addlegendentry{SVR}

            \addplot[thin, fill=brown!20, brown=blue!70, smooth, opacity=0.6] table [x=bin, y=cnt, col sep=comma] {\loadedtableDDDD} \closedcycle;
            \addlegendentry{RF}

            \addplot[thin, fill=red!20, draw=red!50, smooth, opacity=0.6] table [x=bin, y=cnt, col sep=comma] {\loadedtableD} \closedcycle;
            \addlegendentry{Trans}

            \addplot[thin, fill=blue!20, draw=blue!70, smooth, opacity=0.6] table [x=bin, y=cnt, col sep=comma] {\loadedtableDD} \closedcycle;
            \addlegendentry{MLP}

        \end{semilogyaxis}
    \end{tikzpicture}\hspace{-2mm} 
    \begin{tikzpicture}
        \begin{semilogyaxis} [
            log origin=infty,
            title={\tiny Power Consumption}, title style={yshift=-1.5ex},
            ticklabel style={font=\fontsize{4}{5}\selectfont},
            width=0.24\linewidth, height=.25\linewidth,
            xlabel={\tiny relative error (\%)},  x label style={at={(axis description cs:0.5,-0.1)},anchor=north},
            xtick={ 0, .02, .04}, xticklabels = {0, 2, 4}, scaled x ticks=false,
            xmax=0.06, ymin=0, ymax=100,
            ytick={ 10, 50, 100 }, yticklabels={ 10, 50, 100 },
            legend style={at={ (1.0,.6)}, anchor=east, draw=none, fill=none, font=\fontsize{4}{5}\selectfont },
            legend cell align={left}
        ]      

            \addplot[thin, fill=green!20, draw=green!70, smooth, opacity=0.6] table [x=bin, y=cnt, col sep=comma] {\loadedtableEEE} \closedcycle;
            \addlegendentry{SVR}

            \addplot[thin, fill=brown!20, brown=blue!70, smooth, opacity=0.6] table [x=bin, y=cnt, col sep=comma] {\loadedtableEEEE} \closedcycle;
            \addlegendentry{RF}

            \addplot[thin, fill=blue!20, draw=blue!70, smooth, opacity=0.7] table [x=bin, y=cnt, col sep=comma] {\loadedtableEE} \closedcycle;
            \addlegendentry{MLP}

            \addplot[thin, fill=red!20, draw=red!50, smooth, opacity=0.6] table [x=bin, y=cnt, col sep=comma] {\loadedtableE} \closedcycle;
            \addlegendentry{Trans}

        \end{semilogyaxis}
    \end{tikzpicture} 
    \caption{Low-Noise Amplifier}
    \label{fig:error:compare:lna}
    \vspace{-3mm}
\end{figure}

\newpage

\textbf{Results on KNN}

\input{./plot/plot_error_knn}
\end{document}